\documentclass[a4paper,fleqn,usenatbib]{mnras}

\usepackage{newtxtext,newtxmath}

\usepackage[T1]{fontenc}
\usepackage{ae,aecompl}

\usepackage{graphicx}	
\usepackage{amsmath}	
\usepackage{amssymb}	

\newcommand{\mr}{\mathrm}

\usepackage{mathrsfs}
\usepackage{bm}
\usepackage{url}
\graphicspath{{figs/}}
\newcommand{\comm}[1]{}
\usepackage{enumerate}
\usepackage{color}

\title[Cloud signatures in exoplanet transit spectra]{On Signatures of Clouds in Exoplanetary Transit Spectra}
\author[Pinhas \& Madhusudhan]{
Arazi Pinhas$^{1}$\thanks{E-mail: \textcolor{blue}{ap817@ast.cam.ac.uk} (AP); \textcolor{blue}{nmadhu@ast.cam.ac.uk} (NM)} \& 
Nikku Madhusudhan$^{1}$
\\
$^{1}$Institute of Astronomy, University of Cambridge, Madingley Road, CB3 0HA
}

\date{Accepted 2017 July 19. Received 2017 June 29; in original form 2017 February 23}

\pubyear{2017}

\begin{document}
\label{firstpage}
\pagerange{\pageref{firstpage}--\pageref{lastpage}}
\maketitle

\begin{abstract}
Transmission spectra of exoplanetary atmospheres have been used to infer the presence of clouds/hazes. Such inferences are typically based on spectral slopes in the optical deviant from gaseous Rayleigh scattering or low-amplitude spectral features in the infrared. We investigate three observable metrics that could allow constraints on cloud properties from transmission spectra, namely, the optical slope, the uniformity of this slope, and condensate features in the infrared. We derive these metrics using model transmission spectra considering Mie extinction from a wide range of condensate species, particle sizes, and scale heights. Firstly, we investigate possible degeneracies among the cloud properties for an observed slope. We find, for example, that spectra with very steep optical slopes suggest sulphide clouds ({\it e.g.} MnS, ZnS, Na$_2$S) in the atmospheres. Secondly, (non)uniformities in optical slopes provide additional constraints on cloud properties, e.g., MnS, ZnS, TiO$_2$, and Fe$_2$O$_3$ have significantly non-uniform slopes. Thirdly, infrared spectra provide an additional powerful probe into cloud properties, with SiO$_2$, Fe$_2$O$_3$, Mg$_2$SiO$_4$, and MgSiO$_3$ bearing strong infrared features observable with {\it James Webb Space Telescope}. We investigate observed spectra of eight hot Jupiters and discuss their implications. In particular, no single or composite condensate species considered here conforms to the steep and non-uniform optical slope observed for HD~189733b. Our work highlights the importance of the three above metrics to investigate cloud properties in exoplanetary atmospheres using high-precision transmission spectra and detailed cloud models. We make our Mie scattering data for condensates publicly available to the community. 
\end{abstract}
\begin{keywords}
radiative transfer -- scattering -- planets and satellites: atmospheres -- planets and satellites: composition -- planetary systems.\end{keywords}


\section{Introduction}\label{intro}

Transmission spectroscopy has been one of the most successful methods in characterising exoplanetary atmospheres \citep[see e.g.][]{burrows14,madhu16a}. Studies of planets in transit have been used to infer a wide variety of properties such as clouds and hazes, molecular abundances, and pressure-temperature structures \citep{sing16, kreidberg14a, kreidberg14b, knutson14a,knutson14b, madhu14b,pont13, demory13}. One of the major inferences from optical and near infrared (IR) transmission spectroscopy in recent years is the incidence of clouds and hazes in exoplanetary atmospheres. The terms `clouds' and `hazes', collectively regarded as `aerosols', are used in different contexts in the literature. From a formation standpoint, `haze' implies particles formed through photochemical processes whereas a `cloud' constitutes particles formed through condensation of vapour onto a nucleus under suitable thermodynamic conditions \citep{marley13}. On the other hand, these terms are also used in reference to the spectral features they can cause, especially in parametric models used for atmospheric retrieval or otherwise \citep[e.g.,][]{benneke12, kreidberg14b,sing16,macdonald17}. A `cloud' is generally used to mean a source of gray opacity and high optical depth effective below some height in the atmosphere, while a `haze' is represented by an opacity in the optical through a power law dependence on wavelength \citep[see {\it e.g.}][]{macdonald17}. Our work is more closely associated with clouds defined through the formation standpoint, and we therefore use this terminology throughout our work.

Two basic avenues are used to infer the dominant role of clouds in exoplanetary atmospheres. First, observations of transmission spectra with slopes in the optical that deviate from the canonical Rayleigh slope of $-4$ are attributed to the presence of clouds and hazes \citep{pont13,sing16}. Second, nearly flat spectra in the near IR have also suggested the dominance of clouds, especially for Neptune analogues and super Earths \citep{kreidberg14a, knutson14a, knutson14b}.

Cloud inferences are now widely prevalent. For example, \citet{sing16} observed the atmospheres of ten hot Jupiters and interpreted these as ranging from cloudy to clear. Clouds have also been inferred on rocky exoplanets through transmission spectroscopy. The super-Earth, GJ 1214b, has a transit spectrum devoid of absorption features with clouds as the leading explanation \citep{kreidberg14a}. \citet{howe12}'s ad-hoc haze can fit the observations and \citet{morley13} have suggested that KCl and ZnS clouds with large scale heights may cause the flat spectrum if the atmospheric composition is super-solar in metallicity. \citet{mbarek16} have recently extended the putative condensate species responsible for the spectrum of GJ 1214b, raising the possibility of potassium sulfate (K$_2$SO$_4$), zinc oxide (ZnO), and/or graphite clouds. Whilst the strongest interpretation of GJ 1214b's spectrum is clouds, the other possibility is a relatively small atmospheric extent due to a high mean molar mass of atmospheric gas \citep{bean11}. Similar inferences of clouds have been attributed to several other transiting exoplanets with flat spectra such as HD 97658b \citep{knutson14b}, exo-Neptune GJ 436b \citep{knutson14a}, and exo-Uranus GJ 3470b \citep{ehrenreich14}.

Hot Jupiters have been studied in transit more extensively than super-Earths and Neptunian analogues due to more pronounced atmospheric signatures. The transit spectrum of HD 189733b in the optical to NIR bears a steep slope which almost completely obscures Na and K alkali signatures \citep{pont13}; this slope is hypothesised to be due to small (i.e., $\sim $$10^{-2} \mu$m) particulates characterised by efficient isotropic scattering of photons \citep{pont08, etangs08, wakeford15, lee14}. \citet{etangs08} suggest HD 189733b's steep slope to be due to MgSiO$_3$ (enstatite) particulates. \citet{wakeford15} predict clouds dominated by particles of size 0.025 $\mu$m, though their condensate models are statistically poor fits to observations especially for wavelengths below $0.5 \mu$m. \citet{lee16} have recently suggested that the transit spectrum of HD 189733b may be sampling a combination of cloud compositions.

\citet{fortney05} highlights the greater influence clouds have in transit than for normal viewing angle at secondary eclipse. The longer path length offered in transit implies more extinction by cloud particles and conveys their importance in models to be able to relate to transmission observations. To this end, many cloud models have been formulated of which some have been used to understand transit spectra. \citet{ackerman01} present a cloud model for substellar atmospheres which incorporates a balance between upward turbulent diffusion and downward sedimentation. Detailed non-equilibrium dust modeling which includes nucleation, heterogeneous growth, gravitational settling and subsequent evaporation and efficient convection of the evaporated particulates is considered by \citet{helling08} and \citet{helling08b}. \citet{helling16} have used the latter models to investigate cloud physics in HD 209458b and HD 189733b with three-dimensional atmospheric simulations. \citet{wakeford15} have used the analytic model of \citet{etangs08} to explore the transmission spectra of clouds in hot Jupiters. \citet{wakeford16} followed on this with a study of cloud-condensates in super-hot Jupiters, with a focus on cloud formation in WASP-12b. Using a minimal-$\chi^2$ statistical fit, they determine particle sizes of corundum (Al$_2$O$_3$) and perovskite (CaTiO$_3$) in the atmosphere of WASP-12b to be 0.001$\,\mu$m - 0.25$\,\mu$m and 0.025$\,\mu$m - 0.1$\,\mu$m, respectively. \citet{line16} recently investigated the role of inhomogeneous or patchy clouds on transmission spectra of exoplanets. \citet{heng16} associated the sodium and potassium line properties in the optical with the degree of cloudiness in atmospheres of transiting exoplanets. Cloud studies and cloud models such as these and others have been used to explore the diversity of clouds in exoplanetary atmospheres \citep{marley13}. 

Our goal in the present work is to investigate in detail observable cloud spectral features of importance for interpretation of high-precision transit spectra. There are in principle three key observable components to study clouds using transmission spectra: the slope of a spectrum in the optical, the uniformity of this slope, and extinction features in the IR. Whilst gaseous Rayleigh scattering leads to a slope of $-4$, we show that condensates can lead to slopes over a broad range from $-13$ to 1. We study in detail the degeneracies in observed optical slopes that arise from a combination of condensate species, modal particle sizes, and cloud scale heights. We also explore the extent to which these slopes are uniform and thereby their potential to reduce degeneracies in inferring cloud properties. These two observables promise to be powerful means of understanding clouds in the optical, especially with future high-precision observations. We also study condensate signatures in the infrared and show four species to have strong features. This third observable is of great promise due to the imminent launch of the {\it James Webb Space Telescope} ({\it JWST}). The plethora of current and forthcoming observations in the optical and IR domains from {\it Hubble Space Telescope} ({\it HST}), {\it Spitzer}, {\it Very Large Telescope} ({\it VLT}), {\it JWST}, and forthcoming Extremely Large Telescopes (ELTs; e.g., E-ELT, GMT, and TMT) motivate our present study. 

The paper is organised as follows. Our numerical transmission model for clouds is developed in Section \ref{transmission_model}, along with the Mie theory of light interactions with spherical micro-particles. We then use these Mie opacities in Section \ref{results} to address the three key observables discussed above and illustrate their importance for future high-precision observations. In Section \ref{results5}, we discuss applications to hot Jupiters HD 189733b and HD 209458b, which have some of the highest-precision observations. We then discuss model limitations and conclude our work with a review of the essential outcomes of our study in Section \ref{conclusions}.

\section{Methods}\label{transmission_model}
We here develop a model for the transmission spectrum of a cloudy atmosphere. Section \ref{mie_transmission} discusses our numerical transmission model and the slant optical depth of the transit geometry. We then introduce our Mie theory code for a single cloud particle size in Section \ref{mietheory}. We generalise our cloud opacity to include particle size distributions in Section \ref{size_distribution}. Section \ref{free_params} then summarises the model parameters.

\subsection{Transmission Model}\label{mie_transmission}
 We develop a model for the transmission spectrum of a cloudy atmosphere by considering the transit depth. The transit depth is the measured fractional diminution in the stellar light at a given wavelength when the planet transits its host star. Through considering how much flux is absorbed in the planetary atmosphere, we can derive the transit depth as (see Appendix \ref{appendix0})
\begin{equation}
\Delta(\lambda)= \left (\frac{R_{p,\lambda}}{R_{\star}}\right )^2= \frac{2}{ R_{\star}^2}\int_0^{\infty} r(1-e^{-\tau(\lambda,r)}) dr. \label{full_TD}
\end{equation}
Here $\tau(\lambda, r)$ is the slant optical depth along the line-of-sight at a radius $r$ from the planetary centre. The transit depth is the sum of each planetary annulus $2\pi r dr$ weighted by its corresponding absorbance $1-e^{-\tau(\lambda, r)}$, relative to the projected area of the stellar disk. This can be solved exactly numerically. In the present work we assume a simplified numerical model to computing the transit depth. We herein describe our model and its assumptions.

To begin with we consider a simplified approach wherein an effective altitude is used to represent the whole atmosphere. \citet{etangs08} calculated a planet-independent
effective slant optical depth for a range of $R_{p_0}/H$ ($R_{p_0}$ is a fiducial planetary radius at a certain $\lambda$ and $H$ is the atmospheric scale height) such that the translucent atmosphere of a planet produces an equivalent effect as a sharp occulting disk. The sharp occulting disk model has a transit depth of 
\begin{equation}
    \Delta(\lambda)= \left (\frac{R_{p,\lambda}}{R_{\star}}\right )^2= \left (\frac{R_{p_0}+z_{\mr{eff}}(\lambda)}{R_{\star}}\right)^2 \label{approx_TD}
\end{equation}
The slant optical depth at the top of the occulting disk is $\tau_{\mr{eff}}=0.56$ and defines the effective altitude $z_{\mr{eff}}$ at a given $\lambda$. We have carried out a numerical comparison of the exact formulation in Equation (\ref{full_TD}) with the fiducial numerical model in Equation (\ref{approx_TD}) and find they differ only by $\sim$1\%. In our model, the reference pressure $p_0$ associated with $R_{\mr{p_0}}$ is a free parameter and depending on its value, $z_{\mr{eff}}(\lambda)$ can lie either below or above $R_{p_0}$.

The transit depth $\Delta$ or its proxy $z_{\mr{eff}}/H$ computed through our numerical scheme is consistent with that obtained using the formulation of \citet{etangs08}. Both our formulations consider $R_{p_0}$ as a reference level that can vary in the atmosphere through $p_0$. A comparison with a formulation which regards $R_{p_0}$ as a hard surface is discussed in Appendix \ref{appendix0}. In our model, we start our calculation of the slant optical depth $\tau$ from the bottom of an isothermal atmosphere determined by the equilibrium temperature $T_{\mr{eq}}$ at a base pressure of $p_{\mr{base}}=10$ bar. We work differentially upwards assuming spherical symmetry until the effective slant optical depth of $\tau_{\mr{eff}}=0.56$  is reached.

We use a numerical scheme in our work to calculate $z_{\mr{eff}}/H$ as described in Section \ref{z_eff} because the $z_{\mr{eff}}/H$ of \citet{etangs08} cannot be calculated analytically for atmospheres with clouds of scale heights different from the gas scale height and with multiple opacity sources (e.g., clouds, H$_2$ Rayleigh scattering, et al.). Nevertheless, the peculiar combination of clouds and H$_2$ scattering produces slopes which are nearly equivalent to a pure cloud opacity for cloud scale heights larger than $H_c \approx 2H/5$ (the differences being about $\lesssim$1\%). Therefore the analytical formalism of \citet{etangs08} for this combination is valid upwards of $H_c \approx 2H/5$. Within the interval $H_c \approx 2H/5$ to $H_c= H$, the instantaneous slope of the effective altitude at any wavelength is then,
\begin{equation}
 \frac{d(z_{\mr{eff}}/H)}{d\mr{ln}\lambda} \approx \gamma \frac{H_c}{H} \equiv \gamma \aleph_1, \label{dzH_dlnlam}
\end{equation}
where $\gamma$ is the power on the effective extinction cross-section, $\sigma'=\sigma_0 (a)(\lambda/\lambda_0)^{\gamma(a, \lambda)}$ with $a$ the particle size, and $\aleph_1$ is the ratio of the cloud scale height to the bulk atmospheric scale height.  

\subsubsection{Slant Optical Depth}\label{z_eff}
Our numerical model considers various sources of opacity. These are absorption and scattering by cloud condensates, H$_2$ Rayleigh scattering, and absorption by volatile gas species. We consider the latter gas component to include three contributions: H$_2$O, collision-induced absorption (CIA) H$_2$-H$_2$, and CIA H$_2$-He. The important radiative micro-physics is contained in the slant optical depth $\tau(\lambda, y)$, where $y$ is the vertical distance from the base pressure level of the atmosphere $p_{\mr{base}}$. By considering the line-of-sight transit geometry it is possible to derive that the optical depth along the line of sight \citep{fortney05} for a cloudy H$_2$-rich atmosphere is
\begin{align}
\tau(\lambda, y) = & \tau_{\mr{c}}(\lambda, y)+\tau_{\mr{H_2}} (\lambda, y) +\tau_{\mr{gas}} (\lambda, y) \nonumber \\
 \approx &\sigma_{\mr{c}}'(a,\lambda) n_{\mr{tot,b}}\xi_{\mr{grain}}e^{-y/H_c} \sqrt{2 \pi R_{p_0}H_c}+ \nonumber \\
& \sigma_{\mr{H_2}}(\lambda) n_{\mr{tot,b}} \xi_{\mr{H_2}} e^{-y/H} \sqrt{2 \pi R_{p_0}H}+ \nonumber \\ 
& \sigma_{\mr{H_2O}}(\lambda) n_{\mr{tot,b}} \xi_{\mr{H_2O}} e^{-y/H}\sqrt{2 \pi R_{p_0}H}+ \label{opacity_slant} \\
& \sigma_{\mr{H_2-H_2}}(\lambda) (n_{\mr{tot,b}} \xi_{\mr{H_2}})^2 e^{-y/H}\sqrt{2 \pi R_{p_0}H}+ \nonumber \\ 
& \sigma_{\mr{H_2-He}}(\lambda) (n_{\mr{tot,b}})^2 \xi_{\mr{H_2}} \xi_{\mr{He}} e^{-y/H}\sqrt{2 \pi R_{p_0}H}\nonumber , 
\end{align}
where $\sigma_{\mr{H_2}}(\lambda)$ is \citep{seager10}
\begin{equation}
    \sigma_{\mr{H_2}}(\lambda)=\frac{8.14 \times 10^{-57}}{\lambda^4}+\frac{1.28\times10^{-70}}{\lambda^6}+\frac{1.61\times10^{-84}}{\lambda^8}\,[\mr{m^2}].\label{H2_opacity}
\end{equation}
Here, $\sigma'_{\mr{c}}(a,\lambda)$ is the total effective cross-section of the cloud or condensate species; $n_{\mr{tot,b}}$, the total atmospheric number density at the base pressure; $\xi_i$, the mixing ratio of a given opacity species; and $H_c$ is the condensate scale height where we have assumed $n_c(r)=n_{c,b}e^{-y/H_c}$. Finally, we obtain the effective altitude $z_{\mr{eff}}$ through
\begin{equation}
z_{\mr{eff}} = y_{\mr{eff}}-H\mr{ln}\frac{p_{\mr{base}}}{p_0}
\end{equation}
The effective altitude is calculated by determining the height $y$ where a $\tau_{\mr{eff}} = 0.56$ is reached, and subtracting from this the distance between the base location at $p_{\mr{base}}$ and the reference radius $R_{p_0}$ at $p_0$.

\subsection{Mie Extinction from Single Cloud Particles}\label{mietheory}
It is necessary to use a physical theory for the absorption and scattering of light by micro-particles to explore the varied effects of condensates on transmission spectra. Mie theory is such a theory; it is a solution of Maxwell's equations which considers interactions of electromagnetic radiation with spherical particles. When a solid or liquid particle is illuminated by an electromagnetic wave, electric charges in the particle transform into dipolar antennas that re-radiate waves producing `scattered' radiation. The excited elementary charges may also transform a portion of the incident energy into thermal energy in a process of absorption. 

The efficiencies of scattering and absorption as a function of wavelength are quantified by components of a complex index of refraction, $m(\lambda)=n(\lambda)+i\kappa(\lambda)$. The real index of refraction, $n(\lambda)$, informs about scattering whilst the imaginary index, $\kappa(\lambda)$, describes attenuation through absorption. We use the refractive index data from \citet{wakeford15} for condensate species that are expected to condense in hot Jupiter atmospheres and which have experimental data in the optical and infrared: Na$_2$S, MnS, ZnS, MgSiO$_3$, SiO$_2$, Al$_2$O$_3$, FeO, Fe$_2$O$_3$, TiO$_2$, NaCl, KCl, and Fe-rich Mg$_2$SiO$_4$. There are therefore 12 condensate species with 12 elemental compositions represented in our work. We do not consider the effect of rainout of species. In principle, an element in one type of cloud species can limit the formation of other cloud species at higher altitudes due to the depletion of that element. \citet{mbarek16} discuss the effects of rainout on cloud type formation. For example, without including rainout, Fe and FeS clouds may both form in an atmosphere. An account of rainout suggests the availability of Fe undergoes significant depletion so that FeS does not tend to form above the deeper Fe clouds, with S associating to Na rather than Fe to form Na$_2$S condensates at higher altitudes. This can then deplete Na which limits the condensation of NaCl clouds higher in the atmosphere, instead favouring KCl condensation.

\begin{figure*}
    \centering
    \includegraphics[scale=0.55]{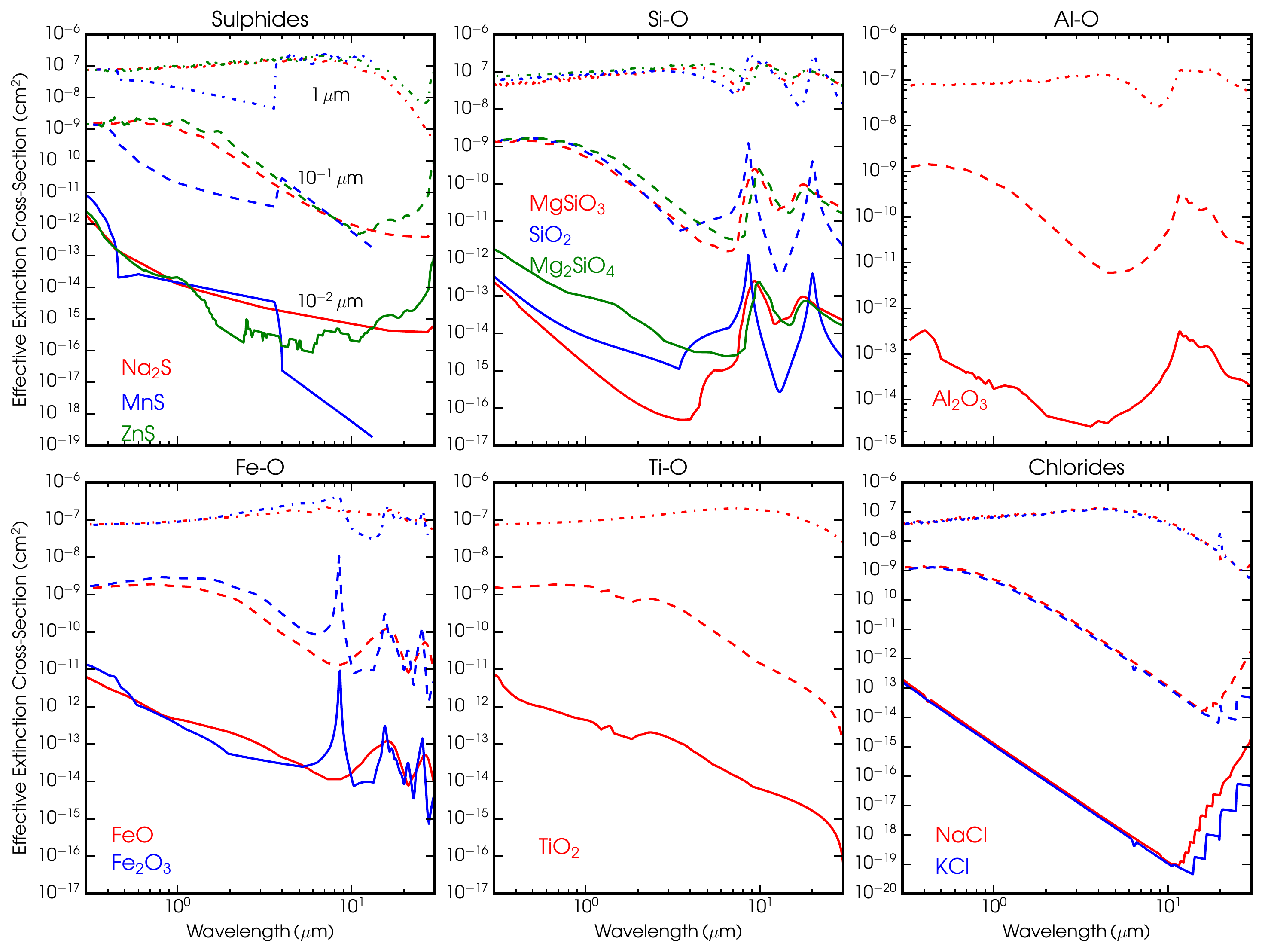}    
    \caption{Effective cross-sections $\sigma'$ of condensates considered in this work calculated using our Mie theory code. Shown are the cross-sections for modal particle sizes of $10^{-2}\,\mu$m (solid), $10^{-1}\,\mu$m (dashed), and $1\,\mu$m (dot-dashed). We consider 12 condensate species with refractive index data from $\sim 0.3\,\mu$m redward, grouped into chemical types.}
    \label{fig:effective_extcs}
\end{figure*}

We implemented our own Mie theory code in Python that is similar to the classical code of \citet{bohren83}. We make our condensate data calculated with our code publicly available to the community.\footnote{We make our Mie theory condensate data publicly available at \url{www.github.com/exo-worlds/Mie_data}} We briefly describe the formulation here. For a spherical particle of radius $a$ embedded in radiation of wavelength $\lambda$, we can define a dimensionless size parameter $x$,
\begin{align}
x &\equiv \frac{2 \pi a}{\lambda}, \,\,\, \mr{where} \, \lambda = \frac{\lambda_0}{m_s}. \\    \label{2.2}
\end{align}
Here, $\lambda_0$ is the wavelength of the incident light in vacuum and $m_s$ is the real refractive index of the ambient medium surrounding the particle. The complex refractive index, $m(\lambda)$, for a particle of specific composition can be used to generate scattering and total extinction cross-sections \citep{deirmendjian69},  
\begin{align}
    \sigma_{\mr{scat}} &= (\pi a^2) \frac{2}{x^2} \sum_{n=1}^{\infty} (2n+1) \{|a_n(m,x)|^2+|b_n(m,x)|^2\}, \\
    \sigma_{\mr{ext}} &= (\pi a^2) \frac{2}{x^2} \sum_{n=1}^{\infty} (2n+1) \, \mr{Re}\{a_n(m,x)+b_n(m,x)\} 
\end{align}
where $a_n$ and $b_n$ are coefficients expressed in terms of Bessel functions of the first kind with fractional orders $\zeta = n \pm 1/2$. The absorption coefficient follows from the relation $ \sigma_{\mr{ext}} =  \sigma_{\mr{abs}} +  \sigma_{\mr{scat}}$. There are multiple ways to express the coefficients $a_n$ and $b_n$, with the most popular in \citet[see page 127]{bohren83} and \citet[see page 17]{deirmendjian69}. Appendix \ref{appendix1} discusses the details of computing these coefficients in more detail. The number of terms needed for computations in the sum for the scattering and extinction cross-sections scales monotonically with the value of the size parameter, and the maximum number of terms needed for good convergence is given by $n_{\mr{max}}=\mr{max} \{x+4x^{1/3}+2, |mx| \}+15$ \citep{bohren83}. 

We use the above formalism to generate the extinction cross-sections for the condensates in our models. In addition to our current application of the theory to exoplanetary atmospheric condensates, Mie theory has been widely used to study the observable effects of grain growth in the context of interstellar and circumstellar environments \citep{draine84, stognienko95, draine06}. In particular, Mie theory and its variants have been recently employed to infer grain size distributions and compositions from spatially resolved sub-millimeter observations of protoplanetary disks \citep[e.g.][]{tazzari16}.

\subsubsection{Effective Cross-Section}

An important feature of scattered irradiation is a strong angular asymmetry characterised by the phase function of scattered radiation. Except for very small particles of $\sim$$10^{-2}\mu$m, scattering ({\it i.e.}, the phase function) of forward-propagating photons is favorably peaked in the forward direction. Indeed, larger particles produce sharper forward peaks. This angular asymmetry is largely due to the fact that radiation from the minute antennas in the forward direction are all in-phase \citep{bohren83}.

Because the measured extinction at an observer is the theoretical  extinction  reduced  by  the scattered light  collected  by  the  detector, the theoretical extinction cross-section is damped by the corresponding asymmetry parameter $g$ characterising the mean cosine of the scattering angle. The effective cross-section can thus be written \citep{hulst80, graaff92}
\begin{equation}
    \sigma' \approx \sigma_{\mr{ext}}-\sigma_{\mr{scat}}g.
\end{equation}
The limiting case of $g=1$ (complete forward-scattering toward the observer) recovers a cross-section due to thermalisation of the grain alone, $\sigma'=\sigma_{\mr{abs}}$. An asymmetry factor of nought is consistent with isotropy, with equal scattering both towards and away from the observer and thus $\sigma'=\sigma_{\mr{abs}}+\sigma_{\mr{scat}}$; scattered light with $g=-1$ is completely obscure to an observer and the total effective cross-section is an enhanced theoretical cross-section, $\sigma'=\sigma_{\mr{ext}}+\sigma_{\mr{scat}}$. 

We show in Figure \ref{fig:effective_extcs} the effective extinction cross-section $\sigma'$ of condensate species for modal particle sizes of $a_0=10^{-2}\,\mu$m, $10^{-1}\,\mu$m, and $1\,\mu$m calculated using our Mie theory code. The mean values of $\sigma_{\mr{ext}}$, $\sigma_{\mr{scat}}$, and $g$ are computed through Equation (13) and Equation (15), respectively. Groups of species have common features in the infrared due to similar chemical bonds and thus similar energy-level excitations. 
The work of \citet{etangs08} and \citet{wakeford15} use the total theoretical extinction cross-section $\sigma_{\mr{ext}}$ as the effective cross-section $\sigma '$ and thereby overestimate the effective altitude $z_{\mr{eff}}(\lambda)$ in transmission spectra, especially for large particle sizes, because the scattered spectrum for large particles is finely distributed around a scattering angle of 0 ($g=1$). Moreover, the slope of the effective altitude is altered because $\sigma_{\mr{scat}}$ and $g$ are both $\lambda$-dependent. Accounting for non-isotropic scattering is crucial and the prescription we use here is a simple correction to include the effect of the angle-integrated scattering field.

\subsection{Grain Abundance}
 The abundance of the grains or aggregates containing the condensed species is $\xi_{\mr{grain}}$. We assume grains to be composed purely of one condensate species. This assumes that homogeneous nucleation is more dominant than heterogeneous nucleation in the formation of condensates, but this may not be true across all atmospheres \citep[cf.][]{marley13}. The grain abundance in Equation (\ref{opacity_slant}) is

\begin{equation}
\xi_{\mr{grain}}=\frac{3 \xi_{\mr{main}} \mu_{\mr{cond}}}{2\rho_{\mr{grain}}\pi a^3}. \label{mixingratio}
\end{equation}
A derivation of this equation can be found in Appendix \ref{appendix2}. Here, $\mu_{\mr{cond}}$ is the mean molecular mass of the condensate species; $\xi_{\mr{main}}$ is the solar abundance of the dominant atomic species in the condensate, for which values are acquired from \citet{burrows99}; and $\rho_{\mr{grain}}$ is the grain density of the species.\footnote{The grain densities are extracted from \url{http://webmineral.com/Alphabetical_Listing.shtml\#\#.V5nTr451o9g}} These chemical characteristics together with the approximate cloud condensation temperatures at $\sim$ 1 mbar are shown in Table \ref{condensate_properties} for the set of 12 condensate species.

\begin{table*}
\begin{center}
\begin{tabular}{||c c c c c||}  
\hline
Condensate Species & Condensation Temperature (K) & $\xi_{\mr{main}}$ & Mean Molar Mass ($\mr{g/mol}$) & Density ($\mr{g\,cm^{-3}}$)\\
\hline

Na$_2$S & 1176 &$1.68 \times 10^{-5}$ & 78.04 & 1.43\\

MnS & 1139 &$1.68 \times 10^{-5}$ & 87 & 4\\

ZnS & 700 &$1.68 \times 10^{-5}$ & 97.45 & 4.05\\

MgSiO$_3$& 1316 & $3.9 \times 10^{-5}$ & 100.33 & 3.2\\

SiO$_2$ & 1725 &$3.26 \times 10^{-5}$ & 60.08 & 2.62\\

Al$_2$O$_3$ & 1677 &$2.77 \times 10^{-6}$ & 101.96 & 4.05\\

FeO & 1650 &$2.94 \times 10^{-5}$ & 71.79 & 5.7\\

Fe$_2$O$_3$& 1566 & $2.94 \times 10^{-5}$ & 159.68 & 5.3\\

TiO$_2$& 1125 & $7.83 \times 10^{-8}$ & 79.86 & 4.25\\

NaCl& 825 & $1.87 \times 10^{-6}$ & 58.44 & 2.17\\

KCl& 740 & $1.23 \times 10^{-7}$ & 74.55 & 2.17\\

Mg$_2$SiO$_4$ (Fe-rich) & 1354 & $3.9 \times 10^{-5}$ & 140.63 & 3.27\\[.5ex]
\hline
\end{tabular}
\caption{Properties of the 12 condensate species considered in this work. The prominent elemental abundances are adopted from \citet{burrows99} whilst the condensation temperatures are obtained from \citet{wakeford15} for a pressure of $10^{-3}$ bar.}
\label{condensate_properties}
\end{center}
\end{table*}

\subsection{Particulate Size Distribution}\label{size_distribution}

The discussion has heretofore assumed a single particle size. However, atmospheres are not composed of one standardised particle size but generally contain condensate grains of many dimensions collectively referred to as polydispersions. We here discuss how we incorporate size distributions into our numerical model for transmission spectra.  

\subsubsection{Mean Parameter Values for Size Distribution}
In order to consider particle size distributions we compute mean values of  the condensate properties. To do so, we weight the scattering and extinction cross-sections, the grain mixing ratio, and the asymmetry parameter by the normalised particle size distribution as follows,
\begin{equation}
\bar{\sigma}_{{\{\mr{ext, scat}\}}}= \frac{\int_{a_1}^{a_2} \sigma_{\{\mr{ext, scat}\}} (a;\lambda) n(a) da}{\int_{a_1}^{a_2} n(a) da},
\end{equation}
and
\begin{equation}
\bar{\xi}_{{\mr{grain}}}= \frac{\int_{a_1}^{a_2} \xi_{\mr{grain}} (a) n(a) da}{\int_{a_1}^{a_2} n(a) da},
\end{equation}
and
\begin{equation}
\bar{g}(\lambda)= \frac{\int_{a_1}^{a_2} g(a;\lambda) n(a) da}{\int_{a_1}^{a_2} n(a) da}, \label{g_sigma}
\end{equation}
where $n(a)$ is the number of particles in unit volume with radii between $a$ and $a+da$ and $g(a;\lambda)$ is the asymmetry parameter for one particle size. Implicit in the above equations is the assumption that the scattering process at one particle is not influenced by other particles. To compute  $g(a;\lambda)$,  there are two elements of the scattering matrix which are important to calculate the intensity of light in any scattered direction. $S_1$ and $S_2$ describe the angular distribution of scattered light assuming unpolarised incident radiation expected from exoplanet host stars. These elements are calculated according to \citet{deirmendjian69}
\begin{align}
S_1 &= \sum_{n=1}^{\infty} \frac{2n+1}{n(n+1)} \{ a_n \pi_n(\cos \theta) + b_n \tau_n (\cos \theta) \}, \\
S_2 &= \sum_{n=1}^{\infty} \frac{2n+1}{n(n+1)} \{ a_n \tau_n(\cos \theta) + b_n \pi_n (\cos \theta) \} 
\end{align}
where $\pi_n (\cos \theta)=P_n(\cos \theta)/\sin(\theta)$ and $\tau_n(\cos \theta)=\frac{d}{d\theta}P_n(\cos \theta)$ are functions of Legendre polynomials $P_n$. These Mie scattering coefficients can also be generated via their own recursion relations without explicit reference to the underlying Legendre polynomials so as to be more useful for our numerical computations. The recursion relations for these angular functions are \citep{bohren83}\footnote{The expression for $\pi_n$ differs from \citet{deirmendjian69}'s by a multiplicative factor of $n$ in the numerator of the third term. In addition, \citet{deirmendjian69}'s expression for $\tau_n(\theta)$ is shown to be in error and has been corrected here.}
\begin{align}
    \pi_n (\theta) &= \cos\theta \frac{2n-1}{n-1}\pi_{n-1}(\theta)-\frac{n}{n-1}\pi_{n-2}(\theta)\\
    \tau_n(\theta) &= n \cos(\theta) \pi_n(\theta) - (n+1) \pi_{n-1} (\theta)
\end{align}
with zeroth, first, and second values as

\begin{equation}
  \begin{split}
    \pi_{-1} (\theta) &=0\\
    \pi_0 (\theta) &=1\\
    \pi_1 (\theta) &=3 \cos(\theta)
  \end{split}
\quad \mr{and} \quad
  \begin{split}
    \tau_{-1} (\theta) &=0\\
    \tau_0 (\theta) &=\cos(\theta)\\
    \tau_1 (\theta) &=3 \cos(2\theta).
  \end{split}
\end{equation}
The scattered intensity in an angular direction $\theta$ characterising the deflection from the original photon direction can be quantified in terms of a probability density function $p(\theta, a)$ called the phase function as \citep{box83} 
\begin{equation}
  p(\theta, a)= 2\pi \lambda ^2 \frac{(|S_1|^2+|S_2|^2)}{\sigma_{\mr{scat}}}.
\end{equation}
The phase function is normalised such that the integral of this PDF over all solid angles $dw=\sin\theta d\theta d\phi$ is $4\pi$. The asymmetry parameter $g(a; \lambda)$ is then incorporated into Equation (\ref{g_sigma}) as
\begin{equation}
  g(a;\lambda)= \frac{\int_{0}^{\pi} \int_{0}^{2\pi} p(\theta; a) \cos(\theta) dw}{\int_{0}^{\pi} \int_{0}^{2\pi}p(\theta; a) dw}.
\end{equation}

\subsubsection{Functional Form of Particle Size Distribution}
Spectra of planetary atmospheres have been fit by various particle size distributions, such as  log-normal distributions, gamma distributions, and power laws. The particle distribution we use is the modified gamma distribution \citep{deirmendjian64} whose general form is
\begin{equation}
    n(a) = \omega a^{\beta}e^{-ba^\alpha}
\end{equation}
and derives its name from the fact that the canonical gamma distribution is recovered for $\alpha \equiv 1$. Here $\omega$, $\beta$, $b$, and $\alpha$ are positive real numbers with the addition that $\alpha$ is an integer. Particular solutions of this general form are infinite and therefore the choice of which to use must involve justification. We use a particular form of the generalised form \citep{budaj15},
\begin{equation}
    n(a) = \left \{ \frac{a}{a_0} \right \}^6 e^{-6a/a_0}
\end{equation}
where $a_0$ is the modal particle size of the distribution. \citet{deirmendjian64} justifies the use of the generalised and specific forms in stating that a survey of the many proposed functional forms shows this particular form of the general modified distribution fits well with measurements of Earth's water clouds and aerosols for $a_0 \approx 4 \mr{\mu m}$, and has the advantage that its parameters have more physical connotations than other distributions. Given the impossibility of {\it in-situ} measurements of cloud particle sizes in exoplanetary atmospheres, the assumed distribution appears a reasonable assumption at present. Given the great variety of nature, however, it is expected that the exact cloud particle size distributions in exoplanetary atmospheres will naturally deviate from that assumed.

\subsection{Free Parameters}\label{free_params}
Our transmission spectrum model contains five free parameters as follows:
\begin{enumerate}[(a)]
    \item Cloud scale height, $H_c$: Most planets in the Solar System including Earth have cloud scale heights $H_c \lesssim H/3$ \citep{fortney05,ackerman01,carlson94,brooke98, lavega04}. Yet dynamical mixing processes in hot Jupiter atmospheres allow considerable deviations from these local findings \citep{parmentier13}, especially for  small grains that do not settle readily. Varying $H_c$ has important ramifications on the transit depth or effective altitude. Decreasing the value of $H_c$ effectively decreases the condensate opacity and hence tends the slope towards the gaseous Rayleigh limit and also lowers the continuum level of the transmission spectrum.
    \item Modal particle size, $a_0$: Particle sizes in planetary atmospheres are expected to be distributed over a continuum of sizes. We assume cloud particles follow a certain distribution with a modal size. Increasing the modal size decreases the slope of the transit spectrum, with a tendency towards flat spectra for the very largest of particle sizes. 
    \item Reference pressure, $p_0$: The pressure associated with the inferred radius $R_{p_0}$ can range from $\sim 1$ bar to $\sim 10^{-3}$ bar. Increasing (decreasing) $p_0$ shifts $R_{p_0}$ lower (higher) in the planet, increasing (decreasing) the transit depth level. 
    \item Grain abundance, $\xi_{\mr{grain}}$: The non-equilibrium nature of cloud condensation makes determining the grain abundance $n_{\mr{c,i}}/n_{\mr{tot,i}}$ difficult and uncertain. We have assumed that the homogeneous grains are limited by the dominant element of the condensate species \citep[cf.][]{etangs08,wakeford15}. 
    \item Molecular abundance, $\xi_j$: The molecular abundances of different volatile species have been suggested to range from subsolar to supersolar \citep{kreidberg14b, madhu14b, macdonald17, fortney13, mordasini16, madhu16a}. We investigate different molecular abundances with cloud types in Section \ref{results4} to determine significant infrared features observable with future {\it JWST} spectra. 
\end{enumerate}

\section{Results}\label{results}

We have developed a model for transmission spectra of cloudy atmospheres. We show a number of its applications to observations following three key observables. We first construct a metric characterising the slope of the transmission spectrum in the $0.3\,\mu$m to $0.56\,\mu$m spectral range. This metric can be useful for constraining cloud compositions, scale heights, and modal particle sizes using high-precision observations. Moreover, we discuss the use of temperature information as a way to reduce degeneracies amongst the different cloud properties. We then continue with a discussion on the uniformity of slopes in the optical as an additional way to break degeneracies using high-precision observations. Finally, we show which condensate features are pronounced in the infrared as a means to identifying dominant cloud signatures with future {\it JWST} spectra.

\subsection{A Metric to Evaluate Optical Slopes}\label{results1}

Transmission spectra of exoplanets in the visible show various slopes that hint towards the existence of different cloud species. We here construct a metric to extract cloud information from the observed spectrum in the optical window of $0.3\,\mu$m to $0.56\,\mu$m. We find this window to be the best metric for the optical slope since it avoids potential contributions from the gaseous sodium and potassium absorption features at approximately $0.59\,\mu$m and $0.77\,\mu$m. It is possible that other species could have spectral features in this range, e.g. TiO and VO \citep{hubeny03}. However, detections of such species are still tentative \citep[see e.g.][]{evans16}, and planets with $T_{eq}$ below 2000 K are less likely to host high temperature gases such as TiO and VO \citep{hubeny03}. We have considered using the slopes between 0.6$\,\mu$m and 0.8$\,\mu$m as a second metric window, neglecting the Na and K features. Using this latter window, however, is not a good predictor of the properties for single species unless the slope in the entire optical is nearly constant.

Observations of hot Jupiter atmospheres obtain the transit depth $\Delta$ from which the total radius of the planet $R_{p_\lambda}$ as a function of wavelength is extracted. The slope of this observed radius is then
\begin{equation}
    m_{\mr{obs}}=\frac{dR_{p_\lambda}}{d\mr{ln}\lambda}=\gamma H_c.
\end{equation}
Here there is a degeneracy in the slope for values of $\gamma$ and $H_c$. Because the cloud scale height $H_c$ is generally difficult to determine {\it a priori}, the slope should be formulated in terms of the more certain bulk atmospheric scale height $H$. The slope of $R_{p_\lambda}=(R_{p_0}+z_{\mr{eff}})$ expressed in a more convenient dimensionless form is 
\begin{align}
     \mathcal{S} \equiv \frac{m_{\mr{obs}}}{H}&=\frac{d(R_{p_\lambda}/H)}{d\mr{ln}\lambda} \nonumber \\
     &=\frac{d(z_{\mr{eff}}/H)}{d\mr{ln}\lambda} = \gamma \frac{H_c}{H},\,\,\,\,\,\text{for $2H/5 \lesssim H_c \leq H$}.\label{normalt_slope}
\end{align}
In the limit of small $H_c$ ($H_c \lesssim 2H/5$), $\mathcal{S}$ tends to $-4.2$ due to Rayleigh scattering from ambient H$_2$. The dimensionless slope $\mathcal{S}$ is determined through dividing the directly observable quantity $m_{\mr{obs}}$ by the atmospheric scale height $H$. The scale height $H$ is typically taken to be the equilibrium scale height, such that $H=H_{\mr{eq}}=k_BT_{\mr{eq}}/(\mu g)$ where the equilibrium temperature obtained from radiative balance is \citep{seager10}
\begin{equation}
    T_{\mr{eq}}=T_{\star}\left(\frac{R_{\star}^2}{2a_{\mr{major}}^2} \right )^{1/4} (f(1-A))^{1/4},
\end{equation}
in which $a_{\mr{major}}$ is the semi-major axis of the planet, $A$ is the Bond albedo, and $f$ is the heat redistribution fraction.
The equilibrium temperature defines an upper bound for the assumed scale height $H$ since the temperature in the terminator region of the atmosphere as observed in transmission spectra may be safely assumed to be smaller than $T_{\mr{eq}}$.

The observer measures a slope $m_{\mr{obs}}$ and divides by an estimate for $H$ to get the slope in the dimensionless form as in Equation (\ref{normalt_slope}). The new slope $\mathcal{S}$ therefore involves a directly observable quantity and an estimated quantity and is thereby sensitive to changes or uncertainty in $H$. The deviation in $\mathcal{S}$ due to an uncertainty in the value $H$ from the actual value is written as
\begin{equation}
    \frac{\delta \mathcal{S}}{\mathcal{S}}=-\frac{\delta H}{H}.
\end{equation}
The uncertainty in bulk atmospheric scale height $H$ enters through uncertainty in the temperature $T$ and the mean molecular weight $\mu$. Assuming knowledge of $\mu$, the primary source of uncertainty is in $T$ such that $-\frac{\delta H}{H}=-\frac{\delta T}{T} = -\frac{T_{\mr{act}}-T_{\mr{es}}}{T_{\mr{es}}}$. The actual temperature in the observable atmosphere at the terminator $T_{\mr{act}}$ is always likely to be less than the estimated temperature $T_{\mr{es}}$ ($=T_{\mr{eq}}$), such that $\delta T$ is negative giving a negative $\delta \mathcal{S}$ for a negative $\mathcal{S}$. For example, if the actual temperature is lower than $T_{\mr{es}} = T_{\mr{eq}}$ by 25 percent ($\delta T/T_{eq}=-0.25$) and $\mathcal{S}=-4$ then the slope can only change by $-1$, i.e. $\delta \mathcal{S}=-1$. Therefore the dimensionless slope $\mathcal{S}$ does not significantly change even for reasonably large $\delta T$. On the other hand, knowledge of $T$ subsumes ignorance into $\mu$, giving instead $-\frac{\delta H}{H}=-\frac{\delta \mu}{\mu}$. Assuming H$_2$O constitutes the only metallic component, the uncertainty in $S$ for an atmosphere of 50 $\times$ solar metallicity for which $H$ is estimated with a solar-metallicity assumption translates to $\delta S = +0.8$. 

\begin{figure*}
    \centering
    \includegraphics[scale=0.65]{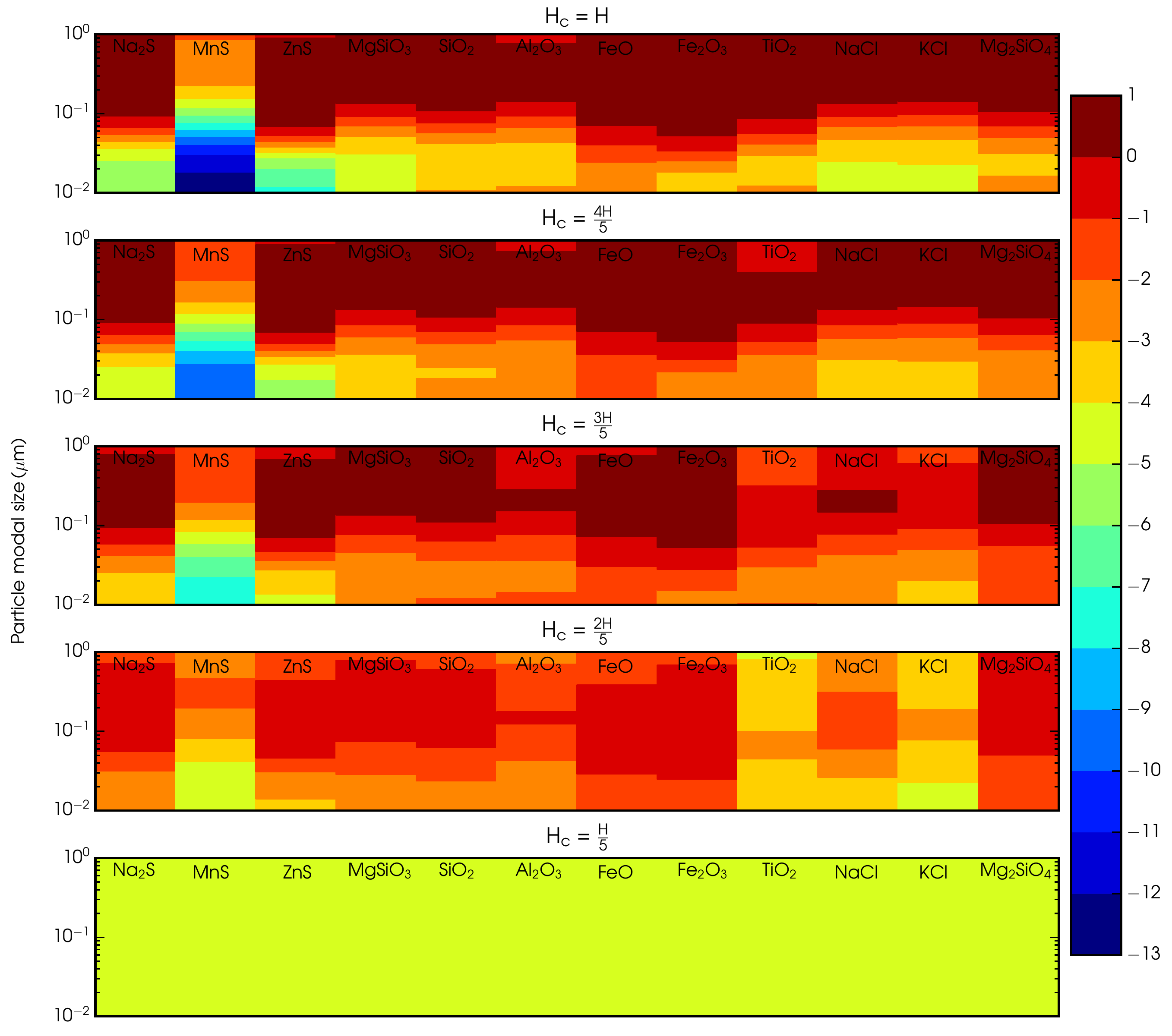}    
    \caption{Best-fitting slopes $\mathcal{S}$ of the transmission spectrum $z_{\mr{eff}}/H$ (see Fig. \ref{fig:zH_slopes1}) in the window 0.3 - 0.56 $\mu$m for the 12 condensate species of various modal sizes and H$_2$ Rayleigh scattering. The transmission spectrum $z_{\mr{eff}}/H$ is calculated using equations (4)-(6). The panels show atmospheres with cloud scale heights of $H_c=H$, $H_c=4H/5$, $H_c=3H/5$, $H_c=2H/5$, and $H_c=H/5$.}
    \label{fig:slopes1}
\end{figure*}

\begin{figure*}
    \centering
    \includegraphics[scale=0.65]{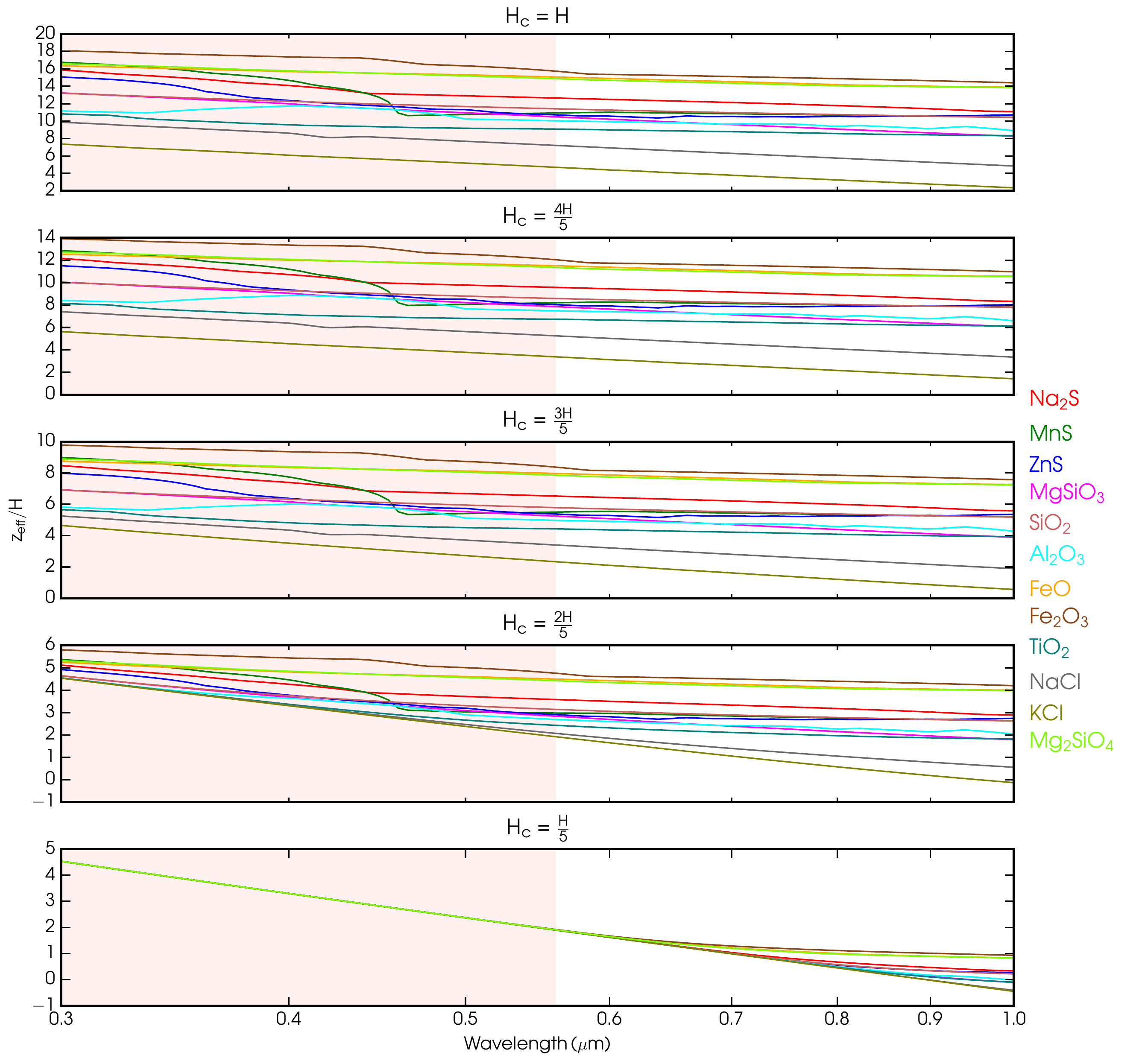}    
    \caption{The transmission spectrum $z_{\mr{eff}}/H$ for the 12 condensate species with $a_0=10^{-2}\mu$m and H$_2$ Rayleigh scattering calculated through equations (4)-(6). The panels show atmospheres with respective cloud scale heights of $H_c=H$, $H_c=4H/5$, $H_c=3H/5$, $H_c=2H/5$, and $H_c=H/5$ that associate with the panels of Figure \ref{fig:slopes1}. The light red area shows the region 0.3 - 0.56 $\mu$m from which the slopes in Figure \ref{fig:slopes1} are calculated.}
    \label{fig:zH_slopes1}
\end{figure*}

In the present work we focus on H$_2$-rich atmospheres which have the most observations. In particular, we focus on transiting hot Jupiters and therefore the mean molar mass of the atmosphere and gravity are taken as fiducial with $\mu=2.3$ g/mol and $g=24.79\,\, \mr{m\,s^{-2}}$. Throughout our work we take the background temperature profile to be an isotherm set to the equilibrium temperature of the planet. We also take the cloud base to be coincident with the bottom of the atmosphere at 10 bar, similar to assumptions from previous studies \citep[c.f.][]{etangs08}. The computed best-fit slopes $\mathcal{S}$ of the transmission spectra in the 0.3 $\mu$m - 0.56 $\mu$m range for the different condensates are shown in Fig. \ref{fig:slopes1}. The slopes are computed for atmospheres of single cloud species together with H$_2$ Rayleigh scattering and for a range of scale heights and modal particle sizes in the modified gamma distribution. Neither variations in $g$, $p_0$, nor $R_{p_0}$ make any difference on the computed slope values. Once $H_c \lesssim 2H/5$ is reached in Figure \ref{fig:slopes1}, the rapid fall of the cloud number density with height means that the H$_2$ Rayleigh scattering slope begins to dominate in the optical. The condensate contribution continues to diminish in progression towards lower $H_c$ and the slope meets a value of $-4.2$ typical of Eq. (\ref{H2_opacity}) at $H_c \approx 0.2 H$. The cloudy models in Fig. \ref{fig:zH_slopes1} show transmission spectra for $a_0 = 10^{-2}\mu$m. The progression $H_c=H\rightarrow H/5$ is made clear, where clouds dominate for large scale heights while H$_2$ Rayleigh scattering dominates for low $H_c$ irrespective of cloud composition. Figure \ref{fig:slopes1} is useful in showing the degeneracy of cloud properties (e.g., composition, modal particle size, and scale height) commensurate with an observed slope. 

There are several important points to be extracted from Fig. \ref{fig:slopes1} as follows:
\begin{enumerate}[(a)]

\item Sulphides are the only species considered over all modal particle sizes and scale heights for which $|\mathcal{S}|>5$. Observations showing steep slopes of $|\mathcal{S}|>5$ suggest sulphide clouds Na$_2$S, MnS, and ZnS.
\item Effects of homogeneous clouds in transmission are observable in the optical for $H_c \gtrsim 0.4H$. Values of $H_c$ less than $0.4H$ suggest two possibilities. $H_c/H \lesssim 0.4$ may suggest the presence of inhomogeneous or patchy clouds (see Section \ref{results5} for an application of this idea to the optical spectrum of HD 209458b) or H$_2$ Rayleigh scattering as the dominant culprit.
\item It is possible to obtain properties of condensate species commensurate with an observed slope when $0.4H \lesssim H_c \leq H$ if we have good knowledge of the $p-T$ profile in the atmosphere. The current state of observations allows for slope degeneracies over cloud composition, scale height, and modal particle size. Large particles (having low $\gamma$, due to a grey $\sigma'$ approached in the geometric limit of Mie theory) with large scale heights are degenerate in slopes with small particles (high $\gamma$, Rayleigh limit) with small scale heights. Unless a more definite calculation for $H_c$ is achieved this latter degeneracy remains present. 
\item A slope of $-4$ does not of itself imply H$_2$ Rayleigh scattering, as such a slope is reproduced by some clouds (e.g., sulphides and chlorides, see Figure \ref{fig:slopes1}) with large scale heights especially in the Rayleigh regime for $a_0\approx 10^{-2} \mu$m (see discussion in Section \ref{results3}).

\end{enumerate}

\subsubsection{Using Temperature to Break Degeneracy}\label{results2}

The temperature structure of the atmosphere can be used as a guide to predict which cloud species are able to condense and thereby reduce degeneracies amongst cloud properties. To first-order, clouds are formed when the temperature of the atmosphere becomes colder than the saturation vapour temperature of a species at a given pressure. The temperature structure can be obtained in the first place by retrieval on observations in primary transit or secondary eclipse \citep{madhu09,madhu11,macdonald17}. Here we use Fig. \ref{fig:slopes1} to illuminate a few important features for the interpretation of transmission spectra assuming knowledge of the atmospheric temperature at $\sim 1$ mbar. We note that using the temperature structure alone neglects other processes that may be active in the atmosphere such as spatial mixing of condensates that can loft deep-forming clouds into the observable atmosphere, complicating a distinct interpretation \citep{parmentier13}. Nevertheless, below we use the cloud condensation temperatures in the observable atmosphere as shown in Table \ref{condensate_properties} to make predictions about possible cloud properties. 

{\textbf{\textit T $\mathbf{\approx}$ 700 K}}:  If the temperature at $\sim 1$ mbar is determined to be less than 700 K from the $p-T$ profile, ZnS (zinc sulfide) and KCl (potassium chloride) may dominate as condensates in the accessible atmosphere. If observations show a slope of $\mathcal{S} < -4$, ZnS clouds may exist in the atmosphere described by modal sizes of $a_0 \approx 10^{-2} \mu$m - $3 \times 10^{-2} \mu$m and $H_c \approx H$ or KCl clouds with sizes $10^{-2} \mu$m - $2\times 10^{-2} \mu$m. As the cloud scale height is decreased to $H_c=4H/5$, slopes decrease correspondingly by 4/5. An important trend which is present for all condensate species is that slopes for large cloud scale heights with large modal particle sizes are mimicked by lower cloud scale heights from smaller modal particle sizes. For example, KCl particles with a modal size of $\sim$ 0.08 $\mu$m assuming $H_c=H$ produce the same slope interval of $-1$ to $-2$ as KCl clouds with $H_c=3H/5$ with modal sizes of 0.05 $\mu$m. This trend follows from a ready understanding of Equation (\ref{normalt_slope}): a large modal size (a low $\gamma$) coupled with a large $H_c/H$ gives the same slope as a small modal size (a high $\gamma$) coupled with a small $H_c/H$. This degeneracy is always present unless there exists a direct method to predict $H_c/H$. A first-order approximation to $H_c/H$ derived from first principles is \citet{lavega04}'s Eq. (22). This equation might be utilized to further constrain the cloud species and modal size responsible for the observed optical slope.

{\textbf{\textit T $\mathbf{\approx}$ 800 K}}: If observations show a slope of $-5 <\mathcal{S} < -3$, NaCl should be well-mixed in the planetary atmosphere and contain grain modal sizes of between $10^{-2} \mu$m and $5 \times 10^{-2} \mu$m. The slopes decrease with gradually decreasing $H_c$ by the appropriate fractions. Scattering by molecular hydrogen begins to occur at $H_c \approx 2H/5$ and a pure Raleigh slope is achieved at $H_c \approx H/5$.

{\textbf{\textit T $\mathbf{\approx}$ 1100 K}}: If $\mathcal{S} \lesssim -8$, MnS (Manganese (II) sulfide) may  exist in the atmosphere for which a direct constraint on the modal size can be determined with the addition that $H_c = H$. On the other hand, there are alternative possibilities if the atmospheric temperatures at $\sim 1$ mbar are thought to be conducive for MnS clouds but observations show slopes of $ -8 \lesssim \mathcal{S} \lesssim -3$. Na$_2$S (sodium sulfide), MnS, and/or TiO$_2$ (titanium dioxide) may exist in the atmosphere. Observations well fit with a straight line leave Na$_2$S as the responsible condensate against MnS and TiO$_2$ (see discussion in Section \ref{results3}). If $\mathcal{S} \gtrsim -3$, there is a degeneracy which arises in that the slopes could be produced either by large particles with $H_c \approx H$ or by small particles with scale heights of $H_c \approx 3H/5$. There is a general pattern in that per given slope, the modal size producing this slope increases sequentially for TiO$_2$, Na$_2$S, and MnS, respectively. Once the scale height becomes lower than $H_c \approx 2H/5$, the magnitude of H$_2$ scattering dominates and the spectrum resembles that of a pure H$_2$ atmosphere by $H_c = H/5$.

{\textbf{\textit T $\mathbf{\approx}$ 1300 K}}: Observed slopes of $\mathcal{S} \lesssim -4$ suggest the existence of MgSiO$_3$ clouds with modal sizes between $10^{-2} \mu$m to $3 \times 10^{-2} \mu$m mixed well with the gas. Mg$_2$SiO$_4$ (forsterite) always has a shallower slope than enstatite for given $H_c$ and $a_0$.

{\textbf{\textit T $\mathbf{\approx}$ 1500 K}}: Slopes of $ -4 \lesssim \mathcal{S} \lesssim -2$ implicate Fe$_2$O$_3$ (iron (III) oxide) with modal sizes $ 10^{-2} \mu$m to $3 \times 10^{-2} \mu$m with $H_c \approx H$. Slopes larger than $-$2 show a degeneracy for different $H_c-a_0$ combinations until $H_c = 2H/5$, at which point the slopes begin to approach a molecular H$_2$ value of $-4$.

{\textbf{\textit T $\mathbf{\approx}$ 1600 K}}: Gradients $\mathcal{S}$ between $-3$ and $-4$ suggest Al$_2$O$_3$ (alumina) rather than FeO (iron oxide) as the condensate species; moreover, these alumina particulates have sizes of about $2\times 10^{-2}\mu$m with a scale height of $H$. Slopes lying between $-2$ and $-3$ may be due to either cloud type with modal sizes of $10^{-2} \mu$m - $2 \times 10^{-2} \mu$m for FeO or $5 \times 10^{-2} \mu$m for alumina, with either option having a scale height of nearly $H$. 

{\textbf{\textit T $\mathbf{\approx}$ 1700 K}}: If $-4 \leq \mathcal{S} \leq -3$, SiO$_2$ (silicon dioxide) may be well-mixed in the atmosphere with modal sizes between $10^{-2} \mu$m and $4 \times 10^{-2} \mu$m, whilst an $\mathcal{S}$ between $-2$ and $-3$ constrains the modal size to $5 \times 10^{-2} \mu$m. The slopes and modal sizes for $H_c= 4H/5$ and $H_c=3H/5$ are almost indistinguishable and remain degenerate unless a physically independent computation of $H_c$ is achieved.

\subsection{The Uniformity of Cloud Optical Slopes}\label{results3}
\begin{figure*}
    \centering
    \includegraphics[scale=0.55]{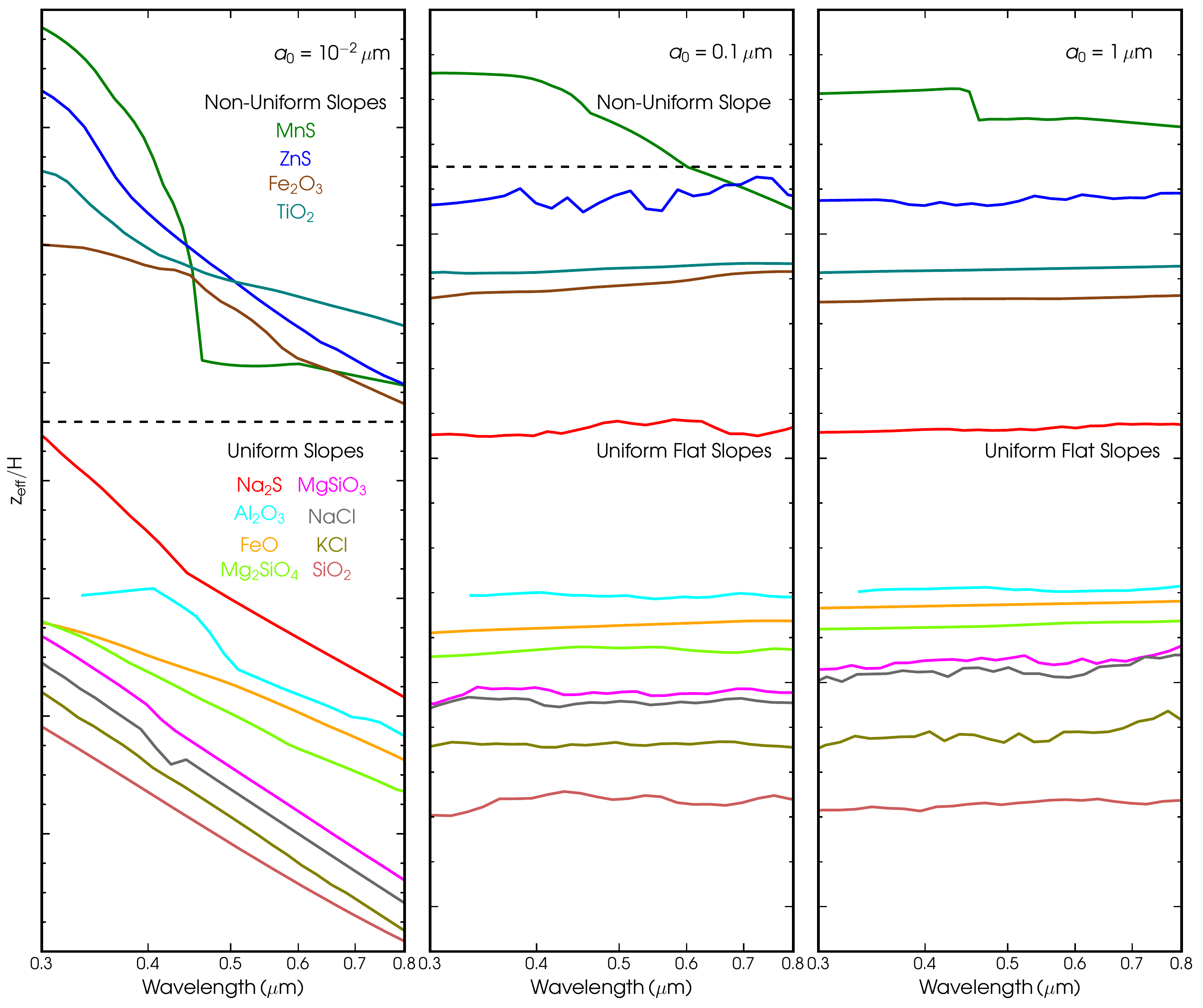}    
    \caption{Model transmission spectra of condensate species showing linear and non-linear trends for three modal particle sizes assuming $H_c=H$. All condensates but MnS, ZnS, Fe$_2$O$_3$, and TiO$_2$ display observationally-limited linear slopes. The dashed black horizontal line delineates between cloud types showing non-uniform and uniform slopes. Vertical offsets have been applied to make these delineations clear.}
    \label{fig:lines_or_not}
\end{figure*}

Most current models of transit spectra use parametric prescriptions for clouds in the optical in both forward models and retrievals resulting in constant slope values \citep[e.g.,][]{sing16}. Observations of current precisions can generally be fit by cloud types with uniform-slope parametric prescriptions. In this section, we explore the uniformity of slopes in the optical for different cloud species. The uniformity of the scattering slopes in the visible provides an additional broad constraint on the condensate composition of special importance for high-precision observations.  

We show the transmission spectra of all 12 condensates with $H_c=H$ for three modal particle sizes ($0.01\mu$m, $0.1\mu$m, and $1\mu$m) in Figure \ref{fig:lines_or_not}. The value $H_c=H$ best highlights the differences between species with non-linear and uniform linear spectra. The condensates fall into two groups according to whether a uniform slope is displayed in the optical range of 0.3 $\mu$m to 0.8 $\mu$m. Our definition of `uniform' is informed by observations. We start at the lowest wavelength and construct a $1-\sigma$ envelope with the same slope as that at the lowest wavelength. A model curve which is able to fit completely within this $1-\sigma$ envelope is considered to have a uniform slope. The average 1$-\sigma$ uncertainty in $z_{\mr{eff}}/H$ for the 10 hot Jupiters in \citet{sing16} is $\sim$1.25 which we use here.

A small number of condensates of modal sizes $\sim$$10^{-2}\mu$m have non-uniform slopes in the optical. As shown in Figure \ref{fig:lines_or_not}, MnS, ZnS, Fe$_2$O$_3$, and TiO$_2$ have the most significant changes in the scattering and absorption properties in the optical leading to large variations in transit spectra slopes. MnS deviates from uniformity most significantly with a broad valley at 0.5 $\mu$m. Fe$_2$O$_3$ shows three distinct regions composing different slopes and TiO$_2$ has a smoother dip than MnS that extends over the whole visible range. On the other hand, high-precision observations showing no significant non-linearity in the visible can reduce species degeneracy to one or more species in the bottom group in Figure \ref{fig:lines_or_not}. Increasing the grain modal particle size to $0.1 \mu$m leads to an essentially grey opacity for all species in the optical except for MnS. For even larger modal particle sizes of $\sim$$1\mu$m, flat spectra result for all cloud species.

In summary, small modal particle sizes of $\sim$$0.01\mu$m produce non-flat spectra. Non-uniform and non-flat spectra are caused by a select few species of MnS, ZnS, Fe$_2$O$_3$, and TiO$_2$. Particle distributions with greater modal sizes of $\gtrsim$$0.1\mu$m produce flat, uniform spectra essentially for all cloud types.

\subsection{Cloud Features in the Infrared}\label{results4}

\begin{figure*}
    \centering
    \includegraphics[scale=0.6]{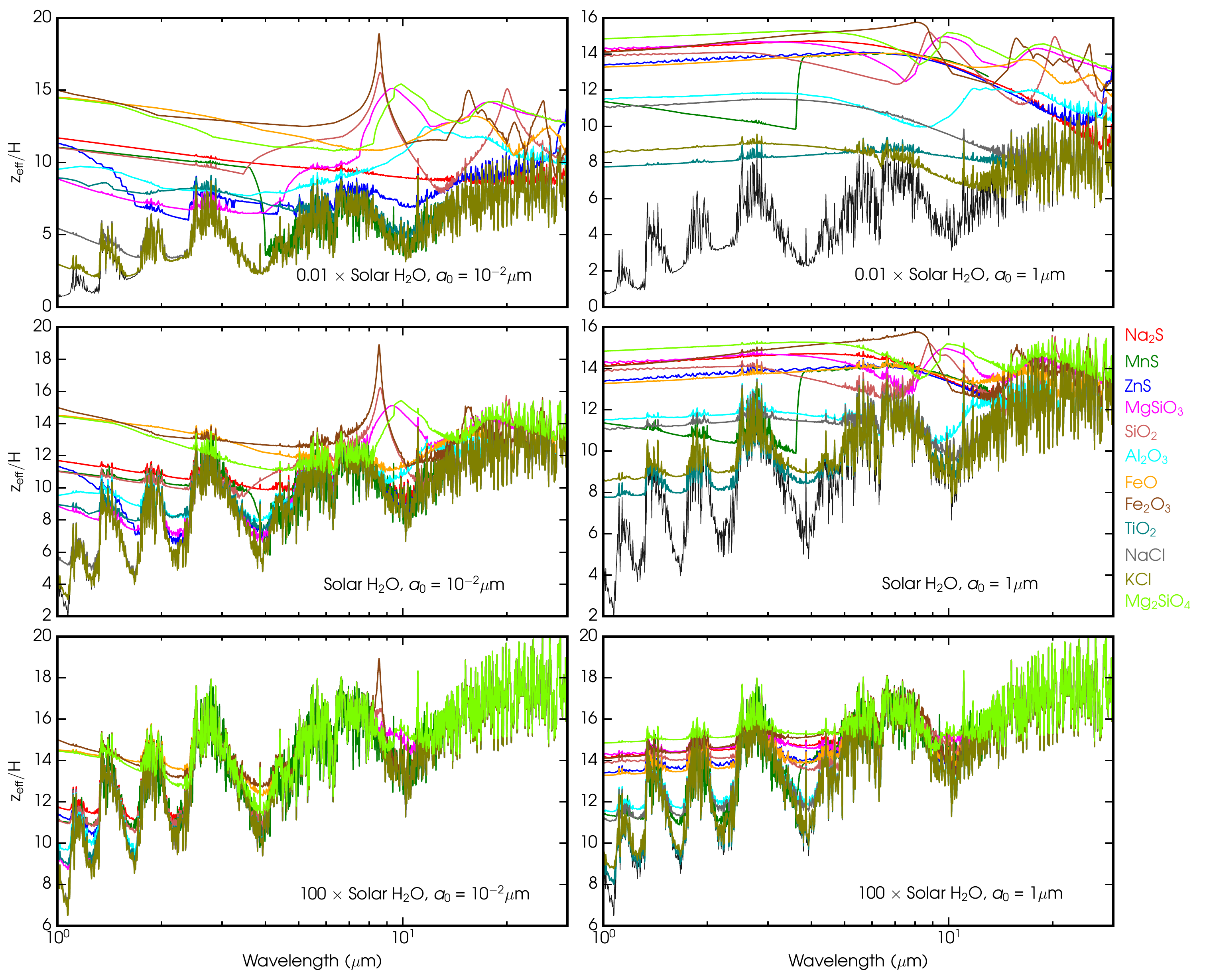}    
    \caption{Model transmission spectra of cloudy hot Jupiter atmospheres with different condensate compositions. The models also contain opacity due to H$_2$O, H$_2$-H$_2$ CIA, and H$_2$-He CIA in the infrared. Each panel represents a particular size distribution and atmospheric H$_2$O abundance and assumes $H_c = H$. The black curves show reference cloud-free models with appropriate H$_2$O  abundances for each panel.}
    \label{fig:clouds_IR}
\end{figure*}

The transit spectrum in the infrared has great potential to reduce degeneracies through identification of condensate signatures. On the other hand, infrared spectra could be dominated by absorption features due to volatile species, e.g. water vapour. We study cloud signatures for different combinations of H$_2$O abundances and modal particle sizes to identify condensate features that are observable in IR spectra, e.g. with {\it JWST}. We generate model transit spectra of cloudy hot Jupiter atmospheres with a representative isothermal temperature of 1450 K.

The abundance of H$_2$O and the modal sizes of cloud particles determine the amplitudes and appearances of condensates features in the infrared. Figure \ref{fig:clouds_IR} shows transmission spectra for six sets of conditions spanning H$_2$O abundances of $10^{-2}\times$ solar, solar, and 100$\times$ solar and $a_0$ of $10^{-2}\mu$m and 1 $\mu$m. These spectra were calculated with additional opacities of CIA H$_2$-H$_2$ and CIA H$_2$-He in Eq. (\ref{opacity_slant}) assuming $H_c = H$. The six combinations illuminate significant condensate signatures that are observable with high precision and high resolution spectra in the infrared. 

\begin{enumerate}[(a)]
    \item {\bf Solar H$_2$O, small $a_0$:} Figure \ref{fig:clouds_IR} shows four cloud species dominate the spectrum for small particulates contained in high H$_2$O abundances at solar value. These species are SiO$_2$, Fe$_2$O$_3$, MgSiO$_3$, and Mg$_2$SiO$_4$. SiO$_2$ and Fe$_2$O$_3$ have strong specral features at $8-9$ $\mu$m, and MgSiO$_3$ and Mg$_2$SiO$_4$ have overlapping absorption signatures at $\sim$$10\mu$m. However, given the lower condensation temperature of Fe$_2$O$_3$ compared to SiO$_2$, distinguishing between these two species may be possible. On the other hand, MgSiO$_3$ and Mg$_2$SiO$_4$ have similar condensation temperatures. Distinguishability between these may come from the amplitude between optical and infrared observations, with MgSiO$_3$ possessing a larger amplitude by a factor of $\sim$4. All four features are of special interest because they occur in a region around $\sim$10 $\mu$m where there is little contribution from H$_2$O absorption, i.e. the valley between the two H$_2$O peaks.
    \item {\bf Subsolar H$_2$O, small $a_0$:} Lower abundances of water vapour with efficiently scattering small particulates bring out more subtle features for the four condensates compared with state (a). SiO$_2$, Fe$_2$O$_3$, MgSiO$_3$, and Mg$_2$SiO$_4$ show multiple features, having additional signatures at wavelengths beyond $\sim$$12\mu$m. MgSiO$_3$ and Mg$_2$SiO$_4$ have overlain features at $\sim$$18\mu$m which may again be distinguished by the difference in optical-infrared amplitudes. An SiO$_2$ peak at $19\mu$m lies in the valley between two prominent Fe$_2$O$_3$ peaks at $\sim$$15\mu$m and $\sim$$22\mu$m, helping to safely distinguish between potential observations in the $\sim$$8\mu$m region. The steep MnS feature at $\sim$$4\mu$m is significant due to its enlargement of the optical-infrared amplitude.
    \item {\bf Solar H$_2$O, high $a_0$:} Large modal particle sizes in solar-abundance atmospheres produce a similar narrative to (a). The main effect of increasing particle dimensions is to transition Mie theory into the geometric limit. The extinction coefficient for large modal sizes approaches a constant value, as seen by the $1 \mu$m curves in Figure \ref{fig:effective_extcs}. The effect is to broaden the narrow features characteristic of small modal sizes. Interestingly, the MnS feature at $\sim$$4\mu$m is inverted relative to case (b). This feature is remarkable because it is not due to absorption as is generally assumed of all condensate features in the infrared \citep{wakeford15}. The scattering and absorption refractive indices for MnS illustrate this feature is due to strong scattering for particle sizes of $\gtrsim 0.1 \mu$m \citep[see MnS panel in Figure 1 of][]{wakeford15}. The case of MnS demonstrates the mistake of this typical assumption, suggesting that scattering may indeed dominate features in the infrared for other cloud species not considered in this work.
    \item {\bf Subsolar H$_2$O, high $a_0$:} Transit spectra of atmospheres with depleted water abundances and large dominant cloud particles are relatively flat with low-amplitude features. As in (c), the amplitude of condensate signatures are quenched with respect to smaller particles but the multiple spectral features of different species in (b) are present. The Al$_2$O$_3$ attribute from $\sim$$10\mu$m$-$$20\mu$m exists only for subsolar water abundances and across all particle sizes. As observed by \citet{wakeford15}, sulphides and chlorides are seen to have no prominent features in the mid-infrared at wavelengths beyond $\sim$$4\mu$m.
    \item {\bf Supersolar H$_2$O:} Highly supersolar (100$\times$) H$_2$O abundances generally supersede all cloud features in the infrared. For small modal particle sizes, the vibrational absorption peak of Fe$_2$O$_3$ at $\sim$9 $\mu$m is still distinguishable from the large-amplitude H$_2$O features. On the other hand, no cloud features are present for large modal sizes. Condensates for which the opacity is large compared to the H$_2$O opacity, e.g. due to large modal particle sizes, exhibit flat spectra in the near infrared.
\end{enumerate} 

As realized by our Figure \ref{fig:clouds_IR} and in \citet{wakeford15}, sulphide and chloride spectral features are not present even in the most promising scenario of subsolar H$_2$O abundances. This allows for spectral interpretation and predictions for planets suggested to host atmospheres with Na$_2$S, KCl, and ZnS condensates, such as GJ 1214b \citep{kreidberg14a} and HD 95678b \citep{knutson14b}. Current observations of GJ 1214b and HD 95678b in transit using {\it HST Wide Field Camera 3} ({\it WFC3}) in the near IR show spectra consistent with being flat. Our work suggests this can be due to KCl condensates of 1 $\mu$m modal particle sizes and with subsolar water abundance. Moreover, Na$_2$S and ZnS can produce flat spectra for 1 $\mu$m modal sizes with subsolar to solar water abundances. {\it JWST} will be capable to observe the atmospheres of these planets at longer wavelengths than 2 $\mu$m. Our Figure \ref{fig:clouds_IR} predicts that future observations of GJ 1214b and HD 95678 should have transit depths which decrease toward longer wavelengths at $\gtrsim$9 $\mu$m if Na$_2$S, KCl, and ZnS clouds indeed dominate the atmospheric spectra.

Our analysis of condensate signatures in the infrared demonstrates four species to be the most conducive for spectroscopic identification with {\it JWST}: SiO$_2$, Fe$_2$O$_3$, MgSiO$_3$, and Mg$_2$SiO$_4$. MnS also has a distinct feature at $4 \mu$m but shows no observable features at longer wavelengths. 
Three instruments aboard {\it JWST} will be important for transmission spectroscopy in the infrared: the Near Infrared Imager and Slitless Spectrograph (NIRISS), the Near Infrared Spectrograph (NIRSpec), and the Mid Infrared Instrument (MIRI). NIRspec can operate in the window 0.6 $\mu$m to 5 $\mu$m in either low or high $R$ ($\lambda/\Delta \lambda$) modes, and will be supplemented with NIRISS from 1.0 $\mu$m to 2.5 $\mu$m with low resolving power. These are complemented with MIRI in the window 5 $\mu$m to 29 $\mu$m with various spectral resolutions $R$, spanning low to high \citep{greene16}.

\section{Application to Current Observations}\label{results5}

\begin{table*}
\begin{center}
\begin{tabular}{||c c c c||}  
\hline
Hot Jupiter & Predicted Condensates & Best-fit slope (0.3 $\mu$m - 0.56 $\mu$m) & $T_{\mr{eq}}$ (K)\\
\hline

WASP-17b & Al$_2$O$_3$&-4.25 $\pm$ 1.03 & 1,740\\

WASP-39b & MnS &-3.89 $\pm$ 1.28 & 1,120\\

HD209458b & MgSiO$_3$ and Mg$_2$SiO$_4$ &-3.04 $\pm$ 0.54 & 1,450\\

HAT-P-1b & MnS &-4.11 $\pm$ 1.73 & 1,320\\

WASP-31b & MgSiO$_3$ and Mg$_2$SiO$_4$ &-5.52 $\pm$ 1.27 & 1,580\\

HAT-P-12b & None &-2.24 $\pm$ 2.88 & 960\\

HD 189733b & MnS & -7.94 $\pm$ 0.61 & 1,200\\

WASP-6b & MnS & -4.14 $\pm$ 1.36 & 1,150\\[.5ex]
\hline
\end{tabular}
\caption{Properties of hot Jupiters considered in this study. Eight hot Jupiters from \citet{sing16} are shown with predicted atmospheric condensates in the observable atmosphere and the best-fit slopes to the transmission spectra in the  0.3 - 0.56 $\mu$m range with associated errors. The expected condensates are obtained from \citet{sing16}'s Figure 2 between the pressures $10^{-1}$ bar and $10^{-3}$ bar for the planetary-averaged $p-T$ profile. The $p-T$ profile along the terminator can vary from this average by $\sim$100 K (see {\it e.g.} \citet{morley16}) but we use these predicted species as fiducial given the overlap in uncertainties in $p-T$ profiles that can be computed from transmission retrieval methods.}
\label{sing16_slopes}
\end{center}
\end{table*}

\begin{figure*}
    \centering
    \includegraphics[scale=0.6]{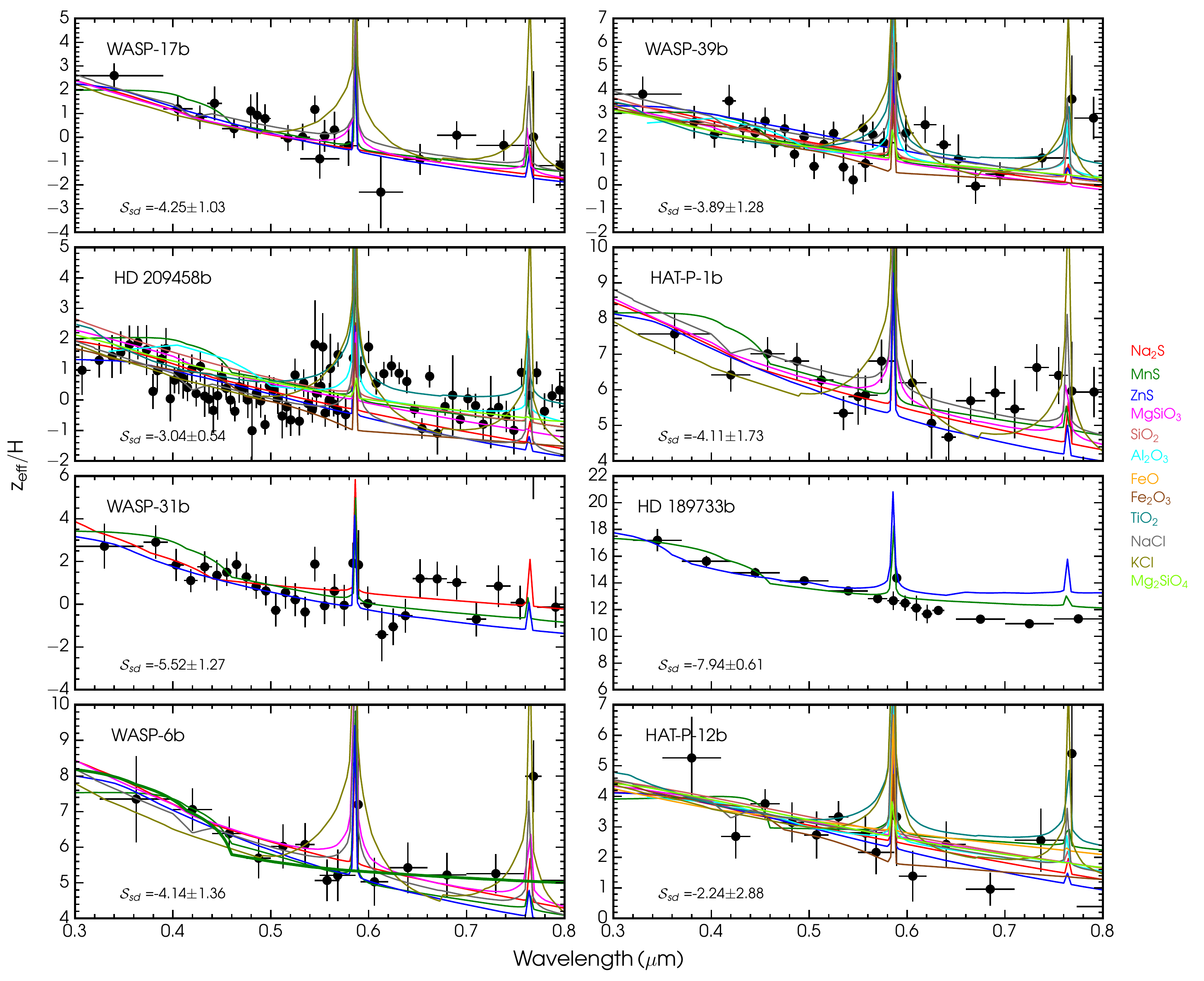}    
    \caption{Forward models of condensate species compared with \citet{sing16} observations for a select eight hot Jupiters. From left to right, top to bottom: WASP-17b, WASP-39b, HD 209458b, HAT-P-1b, WASP-31b, HD 189733b, WASP-6b, and HAT-P-12b. $\mathcal{S}_{\mr{sd}}$ shows the slope of \citet{sing16} data with its associated uncertainty in the window 0.3\,$\mu$m to 0.56\,$\mu$m. The thick green curve in the bottom left panel shows the best-fit model for MnS obtained by minimising the $\chi^2$ statistic.}
    \label{fig:money_plot_optical}
\end{figure*}

We apply the metric from Section \ref{results1} to current observations of eight hot Jupiters. In particular, we discuss two hot Jupiters with the most precise observations, HD 209458b and HD 189733b. For the six other planets we find that current precisions on the spectra allow for a degenerate set of solutions. The latter communicates the importance of improved precisions through multiple-orbit {\it HST} observations. Our models of HD 209458b and HD 189733b illustrate the need for more sophisticated cloudy transit models to interpret current high-precision spectra, as well as the need for higher quality data from future facilities such as {\it JWST} and ELTs. 

Figure \ref{fig:money_plot_optical} shows forward models compared with observations of eight hot Jupiters: WASP-17b, WASP-39b, HD 209458b, HAT-P-1b, WASP-31b, HD 189733b, WASP-6b, and HAT-P-12b. Each panel shows the comparison for one planet and contains the computed best-fit slope and its associated error in the window 0.3 - 0.56 $\mu\mr{m}$ for the observations of \citet{sing16}. \citet{sing16} consider ten hot Jupiters in their study. We do not include WASP-12b and WASP-19b in our study because their atmospheric temperatures are generally hotter than the condensation temperatures for species in Table \ref{condensate_properties}. The suite of models for each planet are calculated assuming $H_c=H$ and the appropriate $a_0$ for each species using Figure \ref{fig:slopes1} that reproduces the central slope values of the \citet{sing16} observations (listed in Table \ref{sing16_slopes}). We assume an isothermal temperature profile valued at the equilibrium temperature of each planet as in Table \ref{sing16_slopes}. We choose $H_c=H$ for our models because condensate species are shown to experience strong mixing in hot Jupiter atmospheres \citep{parmentier13}. We are able to fit the observed spectra for the majority of hot Jupiters as seen in Figure \ref{fig:money_plot_optical}. The observed optical spectra in Figure \ref{fig:money_plot_optical} contain large uncertainties, allowing for degenerate fits to transit observations ranging over different $a_0$ and cloud compositions. In particular, we find multiple indistinguishable fits for WASP-17b, WASP-39b, HAT-P-1b, WASP-31b, WASP-6b, and HAT-P-12b.

We here show the importance of analysing spectra over a broad spectral range from the visible to mid-infrared. As an example, we consider the case of MnS clouds for the transmission spectrum of WASP-6b. Figure 2 of \citet{sing16} shows the planetary-averaged $p-T$ profiles for the eight hot Jupiters along with solar-composition saturation vapour pressure curves. MnS is the only species for which the partial pressure exceeds the saturation vapour pressure in the observable atmosphere of WASP-6b. We compute the minimum $\chi^2$ statistic for MnS by binning the model to the same resolution as the data over the three-dimensional set $\{p_0, H_c, a_0\}$, determining the best parameter set to be $\{0.02\,\mr{bar}, 0.56H, 0.042\,\mu\mr{m}\}$. The bottom left panel of Fig. \ref{fig:money_plot_optical} shows this best-fit model (thick green curve) with a slope of $-4.56$ in the 0.3$\,\mu$m - 0.56$\,\mu$m window, with the best-fit slope of the WASP-6b observations $-4.14 \pm 1.36$ (see Table \ref{sing16_slopes}). Extended into the infrared, however, this minimal $\chi^2$ model is a poor fit with {\it Spitzer Infrared Array Camera} ({\it IRAC}) observations at 3.6$\mu$m and 4.5 $\mu$m. Future cloud models should therefore always account for the entire spectrum simultaneously, especially in retrieval methods. A meaningful cloud statistical fit in one spectral region does not imply fits in other regions.

\begin{figure*}
    \centering
    \includegraphics[scale=0.5]{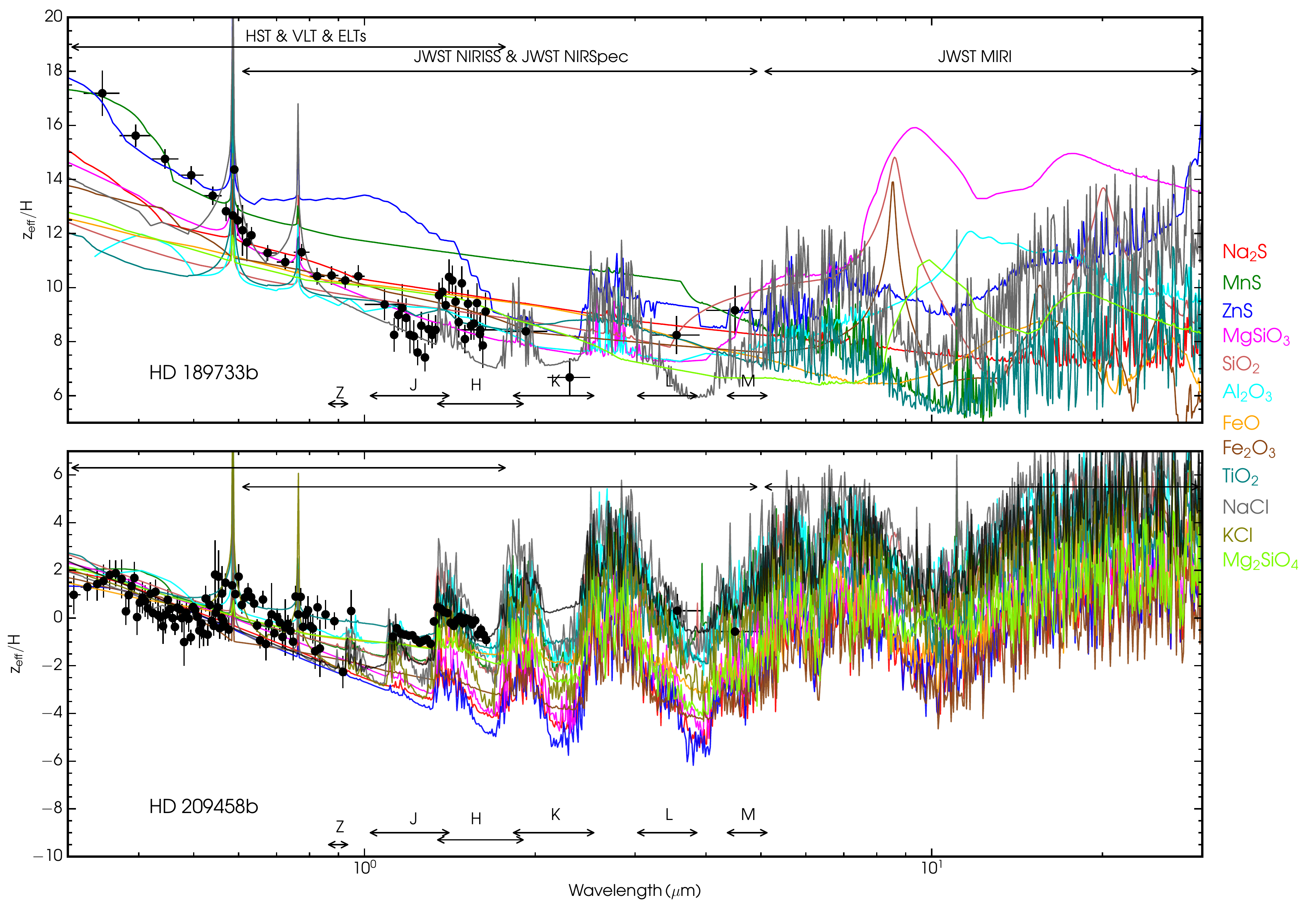}    
    \caption{Transmission spectra of HD 189733b and HD 209458b. Each coloured curve represents an H$_2$-rich atmosphere with the corresponding condensate as shown in the legend, with additional opacities due to H$_2$O, H$_2$-H$_2$ and H$_2$-He CIA, and Na and K at solar abundance. The cloud parameters for each coloured curve are chosen to fit the data using the metric of Fig. 2. HD 189733b and HD 209458b coloured models are for $0.01$$\times$ solar and super-solar H$_2$O abundances, respectively. The black curve is a benchmark model with no condensates nor Na and K opacity, but includes 0.01$\times$solar H$_2$O, and H2-H2 and H2-He CIA opacity. The black circles show data from \citet{sing16}.}
    \label{fig:hd189_hd209}
\end{figure*}

We now focus on two exoplanets with the most precise observations, HD 209458b and HD 189733b. The transmission spectra for HD 189733b and HD 209458b are shown in Fig. \ref{fig:hd189_hd209}.

The observed spectrum of HD 189733b is a challenge to understand since it cannot be reproduced by any single condensate species. Two sulphide species, MnS and ZnS, fit the optical spectrum below 0.6 $\mu$m but slope upwards at longer wavelengths and hence provide a poor fit to the full spectrum overall. A preliminary investigation using combinations of composite species also shows poor fits with the observations. The work of \citet{etangs08} suggests that {\it HST} observations of HD 189733b by \citet{pont08} from 0.55 $\mu$m to 1.05 $\mu$m can be due alone to sub-micron MgSiO$_3$ particulates. On the other hand, \citet{vahidinia14}'s study of HD 189733b illuminates the important effects of  cloud-condensate bases on transmission spectra and suggest the steepness of the spectrum below the observed change in slope at $\lambda \approx 0.6\,\mu$m weakens the role of MgSiO$_3$. Our work finds a similar conclusion to that of \citet{vahidinia14}. We find that consideration of the steepness of the slope at wavelengths below 0.55 $\mu$m disfavors MgSiO$_3$ if the spectrum is assumed to originate from one species alone (see Figure \ref{fig:hd189_hd209}).

This may be a hint to looking at the host star's variability. HD 189733b's host star is active with a significant variation in photometry. Starspots and plages can induce variations in the $z_{\mr{eff}}/H$ optical transmission spectrum \citep{oshagh14,mccullough14}. A combination of plage occultations together with a few cloud-condensate species may well fit the observations. Another possibility is the existence of condensate species that have not been considered thus far.

On the other hand, the spectrum of HD 209458b allows for similar fits in the optical for nearly all cloud species as shown in Fig. \ref{fig:hd189_hd209}. We also include a cloud-free model (black curve) with a water abundance of 0.01 $\times$ solar. The cloud-free water model fits the observations in Figure \ref{fig:hd189_hd209} best, whilst the homogeneous condensate spectra are unable to provide robust fits in the infrared even with substantially supersolar water abundances. The poor fit between homogeneous cloud models and observations lends weight to a patchy cloud scenario, and consideration of the IR region with future {\it JWST} observations has potential to disambiguate the existence of condensate species. We here use a specific idea from our work to substantiate the first suggestion of inhomogeneous clouds on HD 209458b by \citet{macdonald17}. \citet{macdonald17} retrieve the same HD 209458b optical spectrum to obtain $\gamma =  -15.03^{+4.65}_{-3.36}$ and we determine the optical slope to be $\mathcal{S} = -3.04 \pm 0.54$ (see Table \ref{sing16_slopes}). The median cloud scale height with its $1\sigma$ confidence bounds is then $\mathcal{S}/\gamma = H_c/H = 0.20 ^{+0.14}_{-0.06}$. An $H_c$ value of $0.20H$ suggests clouds on HD 209458b may indeed be patchy or inhomogeneous. Our one-dimensional cloud model shows that homogeneous clouds cannot be probed in transmission spectra for $H_c/H \lesssim 0.4$ (see Section \ref{results1}). Given a slope $\mathcal{S}$ that is manifestly less steep than $-4.2$ due to H$_2$ scattering, one possible corollary is that a $H_c/H=0.20$ suggests an azimuthally-averaged value for the day-night terminator, probing inhomogeneous cloud coverage.

The degeneracies and disagreements exposed between our cloud models and present high-precision observations of HD 189733b and HD 209458b call for a detailed approach to cloud modeling in transit spectra. The three key observables using our model are important to interpret cloud properties with high-precision observations using facilities such as {\it HST}, {\it JWST}, {\it VLT}, and ELTs. Of the three key observable cloud properties, condensate features in the infrared show the most immediate promise to cloud characterisation given the imminent {\it JWST} launch. Future infrared observations from {\it JWST} will allow for more detailed studies of clouds with high-precision and high-resolution measurements. Interpretation of the peculiarly steep slope of HD 189733b and HD 209458b's optical data will benefit from {\it JWST}'s broad infrared coverage, enabling study of spectra as a whole from 0.6 $\mu$m to 30 $\mu$m.

\section{Discussions and Conclusions}\label{conclusions}

Using models of transmission spectra of cloudy atmospheres, we illustrate three key observables that serve towards constraining detailed cloud properties with high-precision observations. However, detailed characterisation of current observations require more complicated approaches to modeling clouds in transit. There are many rigorous facets of cloud physics which we have not considered, all of which may need incorporation into more sophisticated models in the future. 

Condensate particles can form either from direct condensation from supersaturated vapour or by condensing on extant  particles of a different composition in processes called homogeneous nucleation and heterogeneous nucleation, respectively. Our model assumes a homogeneous creation of grains, though this assumption may not be the dominant mechanism in some atmospheres \citep{marley13}. We have not modeled the formation mechanism of cloud particulates either through homogeneous or heterogeneous nucleation mechanisms because the simplistic picture of all vapour condensing at saturation ratios above unity is shown to be inaccurate, at least for some combinations of vapours and nucleation seeds. For example, the experiments of \citet{iraci10} suggest that water ice condensation on surrogate materials in the Martian atmosphere is more difficult than presumed. They find that saturation ratios of 2.5 or more are needed for cloud formation. 
    
We have assumed homogeneous, isotropic spherical particles in contrast with irregular particles ({\it e.g.}, ellipsoidals, discs, and fractals). The spherical assumption is idealised and yet has been of great use in understanding many scattering phenomena. We have not considered scattering from arbitrarily irregular particulates for two reasons. First, the domain of validity of modified Mie theories are not extensively tested and are not self-consistent in all cases \citep[see {\it e.g.}][]{schuerman80}. For example, \citet{schuerman80} explains the modified theory of \citet{chylek76} violates energy conservation and predicts total extinction cross-sections that are different from the scattering cross-sections for a purely real refractive index (no absorption), an unphysical result. Second, even under the assumption of self-coherency and exactness, these modified theories are not of much practical use due to our ignorance on the shapes of particles in exoplanetary atmospheres. 

In the present work, we investigate key metrics to characterise clouds in exoplanetary atmospheres using transmission spectra. We construct model spectra to explore three key observables of clouds: the slope in the optical, the uniformity of this slope, and features in the IR. We have explored the first observable through the effects of cloud composition, modal particle size, and scale height on the spectral slope in a clean spectral window in the visible. Second, we have studied which condensates  produce uniform/non-uniform slopes in the optical that are discernible given the precision of current data sets. These two metrics will be of high utility with observations of high-precision transmission spectra in the optical. The third key observable shows the promise of using infrared spectra to further constrain cloud properties.  

Our study of cloud optical slopes shows that very steep slopes of $|\mathcal{S}|>5$ suggest the existence of sulphide clouds. Smaller slope values of $|\mathcal{S}|<5$ show degenerate fits for different cloud species and modal particle sizes for fixed cloud scale height. Consideration of different scale heights still show degenerate fits, in that large modal particle sizes with large scale heights mimic slopes from smaller particles with smaller scale heights. Below a scale height of $H_c \approx 2H/5$ cloud dominance diminishes and the optical slopes of transmission spectra tend towards the H$_2$ Rayleigh scattering value of $-4.2$ in the window $0.3\mu$m$-$$0.56\mu$m. Therefore observable properties of homogeneous clouds are expressed in transmission spectra only for $H_c \gtrsim 2H/5$. Values of $H_c$ smaller than $0.4H$ suggest either inhomogeneous clouds or cloud-free atmospheres dominated by H$_2$ Rayleigh scattering. Finally, a slope of about $-4$ does not of itself suggest H$_2$ Rayleigh scattering since such a slope is produced in the Rayleigh regime for many condensates  for large scale heights (e.g., sulphides and chlorides). One way toward lifting these degenerate fits is through determination of the $p-T$ profile through retrieval methods. 

Observed transmission spectra in the optical are usually fit by straight lines characterized by uniform slopes \citep[e.g.][]{sing16}. However, spectra of condensate species do not all show uniform gradients. For example, whilst NaCl and KCl show linear slopes for a modal size of $0.01 \mu$m, MnS particulates of the same size distribution show a strong characteristic valley at $\sim$0.5 $\mu$m causing significant deviation from linearity. Four of the 12 species considered in this work have significant deviations from uniform optical slopes: MnS, ZnS, Fe$_2$O$_3$, and TiO$_2$. By corollary, the majority of considered cloud types possess observationally-limited uniform slopes in the Rayleigh regime: Na$_2$S, Al$_2$O$_3$, FeO, Mg$_2$SiO$_4$, MgSiO$_3$, NaCl, KCl, SiO$_2$. Future high precision observations in the optical should be able to lift degeneracies in cloud characterisation through these two groups of clouds: those which show substructure in optical slopes and those which are significantly linear. Species can then be further constrained through the first and third observables.

Infrared signatures of cloud species show a promising avenue for discovering condensates in transit spectra. Observations in the infrared have potential to disambiguate degeneracies seen from the optical spectra alone. As against the canonical assumption, we find that infrared features can be due to both absorption {\it and} scattering of radiation. For example, MnS particles of modal sizes $a_0 > 0.1 \mu$m contribute to extinction at $\sim$$4\mu$m by scattering of incident radiation. The cloud features most amenable for future spectral interpretation are those that appear even with high atmospheric water abundances. There are four cloud types with features most promising for future identification with {\it JWST}. SiO$_2$ and Fe$_2$O$_3$ have narrow peaks at $\sim$$8\mu$m$-$$9\mu$m and Mg$_2$SiO$_4$ and MgSiO$_3$ possess broad peaks at broad peak at $\sim$$8\mu$m$-$$12\mu$m. 

We apply our metrics to current observations of eight hot Jupiters. For six of the planets the current precisions on the spectra allow for a wide range of solutions, suggesting the need for higher precision spectra. For the planets with the most precise data, HD 209458b and HD 189733b, we find generally degenerate fits to the optical spectrum of HD 209458b with a few marginally favored solutions, while the spectrum of HD~189733b is challenging to explain by any of the species considered. Overall, our work highlights the importance of broadband (optical-infrared) high-precision transmission spectra as well as detailed theoretical models for reliable inferences of cloud properties in transiting exoplanetary atmospheres. Focused observations with current facilities (e.g. {\it HST} and {\it VLT}) as well as upcoming facilities (e.g. {\it JWST}, ELTs) promise new advancements in this direction. The three empirical metrics presented in our current work will prove useful in the characterisation of clouds in exoplanetary atmospheres using high-precision transmission spectra. 

\section*{Acknowledgements}

AP is grateful for research funding from the Gates Cambridge Trust. AP thanks Rik van Lieshout, Ryan MacDonald, and Siddharth Gandhi for helpful discussions. We thank the referee for useful comments that have added to the quality of our work. This research has made use of NASA's Astrophysics Data System. 



\bibliographystyle{mnras}
\bibliography{references} 

\appendix

\section{Effective Altitude Formulations}\label{appendix0}
There is an alternative formulation to Eq. (\ref{full_TD}) for the effective altitude in transmission derived through considering how much flux is absorbed in the planetary atmosphere. The differential amount of flux traversing the terminator at a radius $r$ from the planetary centre and received by a distant observer at distance $d$ is

\begin{align}
    d\mathcal{F}_{\lambda} \bigg|_{r}&= \mathcal{I}_{\lambda_f} d \mathcal{A}_{r} = \mathcal{I}_{\lambda_f} 2 \pi r dr/ d^2 \\
    &=\mathcal{I}_{\lambda_i}e^{-\tau(\lambda, r)} 2 \pi r dr / d^2
\end{align}
such that the total flux received to the stellar radius (where $\tau \equiv 0$) is thus
\begin{equation}
    \mathcal{F}_{\lambda_r} = \int_0^{R_{\star}} d\mathcal{F}_{\lambda}\bigg|_{r} = \frac{\mathcal{I}_{\lambda_i}}{d^2} \int_0^{R_{\star}} 2 \pi r e^{-\tau(\lambda, r)} dr  
\end{equation}
The difference between the integrated initial stellar flux and the integrated received fluxes, the total absorbed flux, is then
\begin{align}
    \mathcal{F}_{\lambda_a} = \mathcal{F}_{\lambda_i} - \mathcal{F}_{\lambda_r} &= \frac{\mathcal{I}_{\lambda_i}}{d^2} \int_0^{R_{\star}} 2 \pi r dr - \frac{\mathcal{I}_{\lambda_i}}{d^2} \int_0^{R_{\star}} e^{-\tau(\lambda, r)} 2 \pi r dr \\
    \mathcal{I}_{\lambda_i} \pi R^2_{p_\lambda} &= \mathcal{I}_{\lambda_i} \int_0^{R_{\star}} (1-e^{-\tau(\lambda, r)}) 2 \pi r dr
\end{align}
Each planetary annulus is weighted by its corresponding absorbance $1-e^{-\tau(\lambda, r)}$ from the planetary centre outwards, 
\begin{align}
    \pi R_{p_\lambda}^2(\lambda) &=  \pi [R_{p_0}+z_{\mr{eff}}(\lambda)]^2, \label{1stquad} \\
    &= \int_0^{R_{\star}} 2 \pi r (1-e^{-\tau(\lambda, r)})dr \label{2ndquad}
\end{align}
where $R_{p_0}$ is here the radius of the planet for which the atmosphere becomes optically thick to all wavelengths. This formulation is equivalent to that of \citet{dewit13}. Equations (\ref{1stquad} - \ref{2ndquad}) readily become
\begin{equation}
    2R_{p_0}z_{\mr{eff}}(\lambda)+z_{\mr{eff}}^2(\lambda)= \int_{R_{p_0}}^{R_{\star}}2r(1-e^{-\tau(\lambda, r)}) dr.
\end{equation}
This is a quadratic equation for $z_{\mr{eff}}(\lambda)$ whose solution--with the substitution that $z=r-R_{p_0}$ in the integral--gives the final form for the effective altitude,
\begin{equation}
z_{\mr{eff}}(\lambda)=R_{p_0}(\sqrt{1+\aleph_0}-1), \label{zeff}
\end{equation}
where
\begin{equation}
\aleph_0 \equiv \int_0^{R_{\star}-R_{p_0}} \frac{2}{R_{p_0}}\left (\frac{z}{R_{p_0}}+1\right)(1-e^{-\tau(\lambda, z)}) dz.
\end{equation}
As the optical depth $\tau(\lambda, z)$ appears in the exponent and $z$ is always $ \geq 0$, $z_{\mr{eff}} \geq 0 $ and $R_{p}(\lambda) \geq R_{p_0}$. Therefore in this model $R_{p_0}$ acts as a hard surface, whereas in our model it is a reference radius for which $z_{\mr{eff}}$ can lie either below or above.

The instantaneous slope of this effective altitude at any $\lambda$ from Equation (\ref{zeff}) is, 
\begin{equation}
    \frac{d(z_{\mr{eff}}/H)}{d\mr{ln}\lambda}=\gamma \frac{\int_0^{R_{\star}-R_{p_0}}(z/R_{p_0}+1)e^{-\tau}\tau dz}{H(1+\aleph_0)^{1/2}} \equiv \gamma \eta \label{dzH_dlnlam_appendix}
\end{equation}

where $\gamma$ is the power on the effective extinction cross-section, $\sigma'=\sigma_0 (a)(\lambda/\lambda_0)^{\gamma(a, \lambda)}$. This is essentially the equivalent expression to Equation (\ref{dzH_dlnlam}). In the non-Rayleigh limit $\eta$ equals to $H_c/H$ but for small size parameters (or the Rayleigh limit) these two relations are not equal.\footnote{The Rayleigh limit relations for the absorption and scattering cross-sections are formally  applicable in the condition $|m(\lambda)|x<<1$ rather than the oft-cited $x<<1$.} Our numerical formulation of the effective altitude and those of \citet{etangs08} and \citet{wakeford15} use $R_{p_0}$ as a reference altitude for which a negative effective altitude is allowed. This different formulation equivalent to that of \citet{dewit13} is by construction such that $R_{p_0}$ is a hard planetary surface for which the effective altitude can only lie above or at the surface. \citet{betremieux16} treat \citet{etangs08} and \citet{dewit13}'s prescriptions as similar in trying to develop a formalism for the effective altitude which accounts for a hard planetary surface. They are mistaken in both considering the two formulations as similar and moreover in claiming that the model of \citet{dewit13} has no hard `surface'. In fact, as the slant optical depth decreases at the surface (e.g., by considering smaller particles dominating the extinction), \citet{dewit13}'s model {\it does} asymptotically approach the planetary surface without going below, unlike our formulation. \citet{betremieux16}'s Equation (33) is therefore precisely the same as \citet{dewit13}'s; when their $\tau_s \rightarrow 0$, the $\eta \rightarrow 0$ such that the logarithmic slope in Equation (\ref{dzH_dlnlam_appendix}) tends to zero, whilst when their $\tau_s$ tends to large values, the $\eta \rightarrow H_c/H$ such that the logarithmic slope tends to $\gamma H_c/H$. 
\section{Computing Mie Coefficients}\label{appendix1}
We have carried out an extensive study of the \citet{deirmendjian69} form for computing the Mie coefficients $a_n$ and $b_n$. The values for these coefficients are \citet{deirmendjian69},
\begin{align}
    a_n(m,x) =& \{\Theta_1 J_{n+1/2}(x)-J_{n-1/2}(x) \} \times \{ \Theta_1 [J_{n+1/2}(x) \nonumber \\
    & +i(-1)^n J_{-n-1/2}(x)] - [J_{n-1/2}-i(-1)^n \nonumber \\
    &\times J_{-n+1/2}(x)] \}^{-1} \\
    b_n(m,x) =& \{ \Theta_2 J_{n+1/2}(x)-J_{n-1/2}(x) \} \times \{ \Theta_2 [J_{n+1/2}(x) \nonumber \\
    & +i(-1)^n J_{-n-1/2}(x)] - [J_{n-1/2}-i(-1)^n\nonumber \\
    & \times J_{-n+1/2}(x)] \}^{-1}. 
\end{align}
These coefficients are nearly identical except for the differences in $\Theta_1$ and $\Theta_2$ which are $\Theta_1=A_n(mx)/m+n/x$ and $\Theta_2=mA_n(mx)+n/x$ with $A_n(mx)=J_{n-1/2}(mx)/J_{n+1/2}(mx)-n/(mx)$.\footnote{The relations for $a_n$ and $b_n$ in \citet{sharp07} are slightly wrong. We have carried out thorough comparisons of our Mie theory code with the classic work of \citet{deirmendjian69} which reveal this. Equations (33-34) in \citet{sharp07} should have second multiplicative terms with exponents of `-1' instead of the current `1'.}

However, we find that these coefficients are not in forms best suited for computations (see also discussion on page 127 of \citet{bohren83}). For some small volumes of parameter space $\{a, \lambda, n(\lambda), \kappa(\lambda) \}$, the expressions of the \citet{deirmendjian69} formulation break down giving `NaN's. This principally occurs for large parameter sizes $x$. As Section \ref{mietheory} notes, an increased $x$ translates into greater numbers of terms in the sum of the scattering and extinction coefficients. The numerical round-off error associated with finite representation of the irrationally-valued Bessel functions accumulates such that `NaN's arise in evaluations for large number of summation terms. Our Mie theory code therefore follows \citet{bohren83} in computing these two coefficients,
\begin{align}
    a_n(m,x) =& \frac{\{D_n(mx)/m+n/x \} \psi_n(x)-\psi_{n-1}(x)}{\{D_n(mx)/m+n/x \}\xi_n(x)-\xi_{n-1}(x)} \\
    b_n(m,x)=& \frac{\{mD_n(mx)+n/x \} \psi_n(x)-\psi_{n-1}(x)}{\{mD_n(mx)+n/x \}\xi_n(x)-\xi_{n-1}(x)} 
\end{align}
where $D_n= (\mr{ln}\, \psi_n)'$ and satisfies the backward recurrence relation $D_{n-1}=m/(mx)- (D_n+n/(mx))^{-1}$. This recurrence relation is stable when computed from the maximal value of the series $n_{\mr{max}}$ downward. Further, the Ricatti-Bessel functions, $\psi_n$ and $\xi_n$, are
\begin{align}
    \psi_n(z)= &\sqrt{\frac{\pi z}{2}} J_{n+1/2}(z) \\
    \xi_n(z)= &\sqrt{\frac{\pi z}{2}} \{ J_{n+1/2}(z)-iY_{n+1/2}(z) \} = \psi_n(z) - i \chi_n(z)
\end{align}
where $J_{n+1/2}$ and $Y_{n+1/2}$ are the Bessel functions of first and second kind with fractional orders. The Ricatti-Bessel functions satisfy the following recurrence relations computed in ascending fashion
\begin{align}
    \psi_{n+1}(x)= & \frac{2n+1}{x}\psi_n(x)-\psi_{n-1}(x) \\
    \xi_{n+1}(x)= & \frac{2n+1}{x}\xi_n(x)-\xi_{n-1}(x) 
\end{align}
with initial values of 
\begin{equation}
  \begin{split}
    \psi_{-1} &=\cos x\\
    \psi_0 &=\sin x
  \end{split}
\quad \mr{and} \quad
  \begin{split}
    \chi_{-1}  &=-\sin x\\
    \chi_0 &=\cos x.
  \end{split}
\end{equation}

\section{Grain abundance}\label{appendix2}

We here outline the assumptions that go in to the calculation of the condensate grain abundance for a grain composed of a single species. The grain abundance for grains of fixed size assuming a dominant H$_2$ background is 
\begin{equation}
  \xi_{\mr{grain}}=\frac{\mr{Number\,of\,grains\,per\,volume}}{\mr{Number\,of\,H_2\,per\,volume}}=\frac{n_{\mr{grain}}}{n_{\mr{H_2}}}
\end{equation}
The number of grains in unit a volume is
\begin{equation}
  n_{\mr{grain}}=\frac{\mr{Number\,of\,dominant\,atom\,type\,per\,volume}}{\mr{Number\,of\,dominant\,atom\,type\,per\,grain}}=\frac{n_d}{N_d}
\end{equation}
Assuming H$_2$ dominates the number of atoms per unit volume, the number density of the dominant atom type reads

\begin{equation}
  n_{d}= \xi_{d} \times \mr{Number\,of\,atoms\,per\,unit\,volume}= \xi_{d} \times 2 n_{\mr{H_2}}
\end{equation}

As each grain is pure we are enabled to write
\begin{equation}
 N_d = \mr{Number\,of\,condensate\,species\,per\,grain}=\frac{M_{\mr{grain}}}{\mu_{\mr{cond}}}
\end{equation}
With these assumptions, the volumetric number density of the grains is 
\begin{equation}
 n_{\mr{grain}}=\frac{2 n_{\mr{H_2}}\mu_{\mr{cond}}\xi_{d}}{M_{\mr{grain}}}
\end{equation}
A spherical grain implies $M_{\mr{grain}}=4\pi a^3 \rho_{\mr{grain}}/3$ and therefore the averaged grain abundance is
\begin{equation}
\xi_{\mr{grain}} = \frac{3 \xi_{d} \mu_{\mr{cond}}}{2\rho_{\mr{grain}}\pi a^3}.
\end{equation}


\bsp	
\label{lastpage}
\end{document}